\documentclass[aps,prb,twocolumn,nobalancelastpage,nobibnotes,nofootinbib,longbibliography,superscriptaddress]{revtex4-2}
\usepackage{mathrsfs}
\usepackage{amsmath,textcomp,gensymb}
\usepackage{amsfonts}
\usepackage{amssymb}
\usepackage{amsthm}
\usepackage{graphicx}
\usepackage{natbib}
\usepackage{color}
\usepackage[colorlinks=true,citecolor=blue,linkcolor=blue, urlcolor=cyan]{hyperref} 
\usepackage{bm}
\usepackage[caption=false]{subfig}
\usepackage{verbatim}

\usepackage[table, dvipsnames]{xcolor}
\usepackage{cancel}
\usepackage[normalem]{ulem}

\usepackage{empheq}
\definecolor{myblue}{rgb}{.8, .8, 1} 
    
\usepackage[bbgreekl]{mathbbol}

\newcommand{\rot}{\mathop{\rm rot}\nolimits}
\newcommand{\divv}{\mathop{\rm div}\nolimits}
\newcommand{\grad}{\mathop{\rm grad}\nolimits}

\newcommand{\A}{\Tilde{\mathcal{A}}_0}

\usepackage{datetime}

\newcommand{\addZakhar}[1]{\textcolor{black}{#1}}

\begin{document}
\title{Boltzmann transport theory of magnon-exciton drag}

\author{Zakhar A. Iakovlev}
\affiliation{Ioffe Institute, 194021, St. Petersburg, Russia}

\author{Akashdeep Kamra}
\affiliation{Department of Physics and State Research Center OPTIMAS, Rheinland-Pf\"alzische Technische Universit\"at Kaiserslautern-Landau, 67663 Kaiserslautern, Germany} 

\author{Mikhail M. Glazov}
\affiliation{Ioffe Institute, 194021, St. Petersburg, Russia}

\begin{abstract}
We develop a microscopic theory of magnon-exciton drag effect in a bilayer van der Waals antiferromagnetic semiconductor CrSBr. Effective exciton-magnon coupling arises from an orbital mechanism: Magnons tilt the layer magnetizations, enabling charge carrier tunneling that mixes intra- and interlayer excitons and thereby modulates the exciton energy. \addZakhar{We derive the effective Hamiltonian of exciton-magnon coupling, based on our calculation of the magnon spectrum taking into account short-range exchange interaction between Cr-ion spins, single-ion anisotropy, and long-range dipole-dipole interactions.} 
We show that despite rather small renormalization of exciton's energy and effective mass by the exciton-magnon interaction, the three key two-magnon processes: exciton-magnon scattering, two-magnon absorption by an exciton, and two-magnon emission are highly efficient. 
By solving the Boltzmann kinetic equation, we evaluate a short exciton-magnon scattering time which is in the sub-ps range and strongly decreases with the increase in magnon population. Hence, exciton-magnon scattering is likely to be dominant over other scattering processes related to exciton-phonon and exciton-disorder interactions. We demonstrate that \addZakhar{non-equilibrium} magnons can efficiently drag excitons, resulting in a large and nearly isotropic exciton propagation that can significantly exceed the intrinsic anisotropic diffusion.
Our results provide a theoretical basis for recent observations of anomalous exciton transport in CrSBr [\href{https://www.nature.com/articles/s41565-025-02068-y}{F. Dirnberger, et al., Nat. Nano. 21, 65-70 (2026)}] and establish magnon-exciton drag as a powerful mechanism for controlling exciton propagation in magnetic systems.
\end{abstract}

\maketitle


\section{Introduction}\label{sec:intro}

The recent emergence of two-dimensional van der Waals magnetic semiconductors~\cite{Wang:2022,Burch:2018,Brennan:2024} has enabled a new fertile playground concurrently hosting mutually coupled magnetic order and excitons, electron-hole pairs bound by the Coulomb forces. Due to the varied nature of the magnetic order and its spin excitations as well as excitons in this class of materials~\cite{Dirnberger:2022,Wang:2022}, numerous phenomena emerging from the mutual interaction between these excitations are rapidly being discovered. A prominent role is being played by CrSBr due to its hosting robust magnetic order up to a N\'eel temperature of about $130$~K and bright excitons with large binding energy and a substantial Wannier character~\cite{Brennan:2024,Wilson:2021aa}. Magnetically, CrSBr is an A-type two-sublattice antiferromagnet hosting ferromagnetic alignment within each layer, which belongs to one magnetic sublattice, and relatively weak antiferromagnetic interlayer exchange. Consequently, its magnetic ground state can be tuned from antiparallel alignment between the two sublattice magnetizations at zero applied magnetic field to parallel alignment using external fields of the order of 1 T. This evolution of the angle between the layer magnetizations further controls, via spin-dependent interlayer hopping, the ability of the charge carriers to move between the adjacent layers and thus the energy of the excitons~\cite{Brennan:2024,Wilson:2021aa,PhysRevB.111.075107,Semina:2025}. Hence, a strong dependence of the exciton energy on the angle between the two sublattice magnetizations~-- the canting angle, controlled via the applied magnetic field, has served as the first manifestation of the exciton-magnon coupling in CrSBr~\cite{Wilson:2021aa}.

Among the range of possibilities enabled by this exciton-magnon coupling, we note the spatio-temporally resolved detection of exciton energy as a means to observe the corresponding magnetization dynamics. This includes the situation when coherent spin waves or magnons are launched by an ultrafast optical pump pulse and cause a modulation of the sublattice canting angle~\cite{Bae:2022,Diederich:2022,Dirnberger:2023}. Furthermore, the dependence of this average canting angle on the population of incoherent thermal magnons allows sensing of the latter via the exciton energy~\cite{Dirnberger:2023}. Going forward, the reciprocal effect in which the exciton population affects the magnetic canting has also been found to underlie excitonic interactions and nonlinearities~\cite{Johansen2019,Datta:2025,Han:2025}. Several of these effects follow the same phenomenology as their counterparts emerging from the phonon-exciton coupling~\cite{Yazdani:2023}, with the added feature of a convenient magnetic control over the exciton-boson coupling. Consequently, analogues of phonon-drag and \mbox{-wind} effects~\cite{PhysRevB.100.045426} on exciton transport can be anticipated due to magnons, and have been observed recently~\cite{crsbr:exp}.  

\begin{figure}
    \centering
    \includegraphics[width=\linewidth]{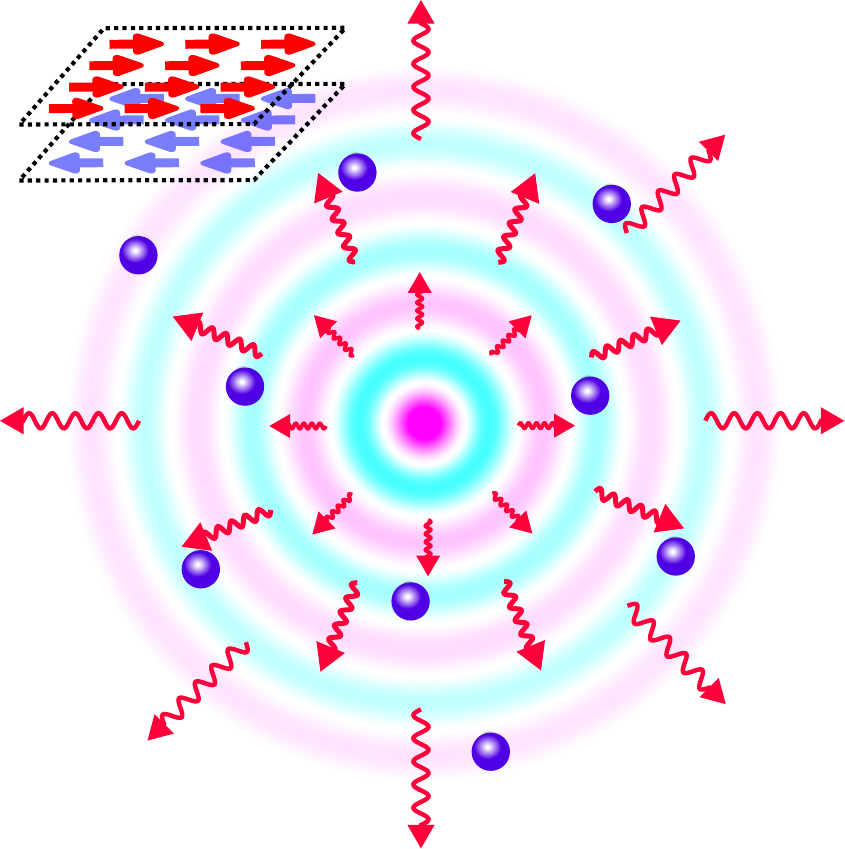}
    \caption{Schematic illustration of magnon-exciton drag effect. Magnons (wavy red arrows) propagate out of the excitation spot and drag excitons (blue balls). Inset shows antiferromagnet bilayer under study.}
    \label{fig:drag}
\end{figure}

The drag phenomenon refers to the motion of one species of particles contributing to the flow of the other due to momentum-conserving scattering between them. It results from a momentum transfer and is ubiquitous. Magnon-electron drag~\cite{Bailyn:1962} was first observed in an antiferromagnetic semiconductor MnTe as an enhanced thermoelectric power around the N\'eel temperature $T_N$~\cite{Zanmarchi:1968}. This enhancement, which manifests itself as a theoretical divergence around $T_N$ in a magnon-based theory~\cite{Sugihara:1973}, is attributed to an increase in the spin excitations' population as one approaches $T_N$. Considering the strongly bound nature of the excitons in CrSBr, one may then expect the emergence of a similar magnon-exciton drag in CrSBr, which has been reported recently~\cite{crsbr:exp}. The key experimental signature of this magnon-exciton drag effect has been a large and nearly isotropic exciton diffusivity, which peaks at $T_N$, despite the highly anisotropy nature of the exciton effective mass in CrSBr~\cite{Brennan:2024,Klein:2023}. Despite the phenomenological similarity between magnon-electron and magnon-exciton drags, a crucial difference arises as excitons, unlike conduction electrons, require a large energy ($\sim 1$ eV per exciton) to be excited and are bound to be less dense than the magnons under a wide range of conditions. Consequently, magnons are expected to drag excitons along due to the former's higher population while the reverse effect is expected to be weak, see Fig.~\ref{fig:drag}. 

Another feature that makes magnons unique  is the long-range magnetic dipole-dipole interactions that are known to result in the GHz-frequency magnetostatic modes~\cite{Damon:1961,Kalinikos:1986} with intriguing properties such as negative dispersion for low wavevectors. These have been studied in great detail for ferromagnets, but are also found in antiferromagnets~\cite{Stamps:1984,Kamra:2017,Kanj:2023}. Recently, some of the key features observed in ultrafast pump-probe experiments in CrSBr regarding excitonic detection of coherent magnons have been interpreted in terms of these magnetostatic modes suggesting a crucial role for the consideration of the dipolar interactions~\cite{Sun:2024aa}. Other works have attributed these observations to an interplay between launching of coherent phonons by the pump pulse and a subsequent magnetization dynamics due to the magnetoelastic coupling~\cite{Bae:2022,Ranhili:2025}, disregarding dipolar interactions. Hence, a detailed theoretical understanding of exciton-magnon coupling and the consequent phenomena duly accounting for the dipolar interactions is highly desirable~\cite{Liu:2024}, which is one aim of the present work.    

In this article, we present a Boltzmann theory of the coupled exciton and magnon transport in a bilayer of CrSBr with a focus on the magnon-exciton drag effect~\cite{crsbr:exp}. We also provide a microscopic model of the exciton-magnon coupling that underlies this drag, and other previously observed effects. In our pursuit of an analytic theory and understanding the key qualitative phenomena, we restrict our considerations to temperatures not too close to $T_N$, low exciton densities, and quasiequilibrium distributions of magnons and excitons. Nevertheless, our analysis and results allow us to comment on a broader operation regime and offers a starting point for its detailed investigation in the future.

\section{Summary of key results}\label{sec:summary}

We begin by presenting the structure of the paper together with a physical intuition-oriented summary of our key results. The discussion in this section is aimed at readers who are not interested in the particularities of theoretical framework and mathematical details.

With the ultimate goal of describing the transport of mutually coupled excitons and magnons in a CrSBr bilayer, we first evaluate its magnon dispersion in Sec.~\ref{sec:magnon}. Here, our system consists of two layers of antiparallel oriented magnetizations in the ground state, with spins within each layer aligned parallel to each other. In evaluating the magnon dispersion, we consider a discrete lattice model accounting for short-range intralayer ferromagnetic exchange, interlayer antiferromagnetic exchange, single-ion anisotropies as well as long-range dipolar interactions between the magnetic moments. While the dipolar fields in the ground state of a collinear antiferromagnet vanish due to the zero net magnetic moment, they play an important role for magnons with small wavevectors because of the nonzero spin, and thus magnetic moment, of the magnons. In agreement with previous works~\cite{Scheie:2022aa,Bae:2022,Sun:2024aa}, the typical slightly anisotropic dispersion for CrSBr magnons dominated by the exchange interaction is obtained at not too small wavevectors. A key effect of the dipolar interactions is the emergence of negative group velocity for one of the magnon branches at low wavevectors. In Appendices \ref{app:semiclass} and \ref{app:FiniteThickness}, we evaluate the magnon dispersion for CrSBr thin films with finite thickness and find that the negative group velocity increases in magnitude with the film thickness, consistent with other recent results~\cite{Sun:2024aa}. In that case, we make use of the alternative yet equivalent approach based on the accounting for the macroscopic magnetic fields produced by the magnon magnetization. Overall, this emergence of negative group velocity and effects of dipolar interaction are similar to the corresponding phenomena studied previously for ferromagnetic and antiferromagnetic thin films. They can be understood in terms of the anisotropic nature of the dipolar interactions resulting in magnon energies depending on the wavevector, which includes the out-of-plane variation in the magnon profile. 

Next, in Sec.~\ref{sec:coupling}, we derive the exciton-magnon coupling for the predominantly intralayer excitons that have been widely studied in this material. Our quantum treatment here of the exciton-magnon coupling complements and offers new insights on the previous phenomenological semiclassical frameworks~\cite{Bae:2022,Dirnberger:2023}. Considering that a noncollinearity between the magnetizations within the two layers allows the interlayer hopping of charge carriers, otherwise forbidden by spin conservation, the intralayer excitons acquire a small interlayer component due to the noncollinearity~\cite{Wilson:2021aa,PhysRevB.111.075107,Semina:2025}. Consequently, the exciton energy depends on the latter. Expressing this noncollinearity in terms of the magnon modes, we obtain the desired Hamiltonian characterizing the exciton-magnon coupling.

Having evaluated this mutual interaction, we examine the effect of magnon cloud on excitonic quasiparticles and their bandstructure via the formation of exciton-magnon polaron~\footnote{This should be contrasted with and is very different from a magnon-polaron, which describes a quasiparticle formed from resonant hybridization between a magnon and a phonon~\cite{Kamra2015,Kikkawa:2016}. Here, the exciton-magnon polaron is a predominantly excitonic excitation that is slightly and non-resonantly dressed by magnons. It is analogous to a polaron~\cite{doi:10.1080/00018735400101213}, which is an electron that gets non-resonantly dressed by lattice distortions.} in Sec.~\ref{sec:polaron} relegating certain theoretical details to Appendix \ref{Appendix:XMP}. We find that this effect is small, causing less than $1\%$ change in the exciton effective mass, and does not significantly influence exciton transport. 

We formulate the Boltzmann transport description for the coupled exciton and magnon clouds in Sec.~\ref{sec:transport}. We work in the small exciton population limit, valid in typical experiments with not too high pump fluences, while the magnon population is assumed to be relatively large. This hierarchy is justified by the fact that both thermal fluctuation energies at reasonable temperatures and magnon energies cover a range from sub meV to few meVs, while the exciton energies are in eVs and all excitons are created by light. Furthermore, we model the nonequilibrium population and transport via quasiequilibrium distribution functions for both excitons and magnons. Within this approximation, the low wavevector and negatively dispersing magnon modes do not play a significant role in the coupled transport. Therefore, the dipolar interactions are not central to the physics here. Thus, the linear regime investigated here contrasts with and is complementary to the consideration of far-from-equilibrium distributions and transport in Ref.~\cite{crsbr:exp} although it already captures the key physics involved. 

Based on our derived exciton-magnon coupling, we show that only four-particle processes are allowed. Either an exciton scatters with a magnon or it absorbs/emits two magnons, bringing the total number of quasiparticles going into and out of the process to 4. Relegating the details to Appendix \ref{app:collision}, we evaluate the rates of these processes and thus the collision integral in the formulated Boltzmann equation. These rates are found to be in the sub-picosecond range. This calculation underscores the critical importance of exciton-magnon coupling for exciton transport in this material. A key reason for these large rates is their inverse dependence on the magnon anisotropy gap, which is typically very small and is in the $\mu$eV or GHz range. This dependence, in turn, appears to stem from the 2D nature of the system and is reminiscent of the infrared catastrophe that is known to impede order in low-dimensional systems. 

We present our key results on the magnon-exciton drag in Sec.~\ref{sec:drag}. Solving the transport equations, we find that the drift velocity of excitons becomes proportional to that of the magnons signifying the magnon-exciton drag effect. Further, the magnon-exciton scattering can be large enough such that the exciton velocity approaches its magnon counterpart, which is the maximum allowed value under our assumption of quasiequilibrium distributions. We also find that the effective diffusivity 
of excitons acquires a nearly isotropic contribution due to the magnon-exciton drag and nearly isotropic magnon dispersion. This contribution can become much larger than the intrinsic diffusivity of the excitons, which is highly anisotropic~\cite{Klein:2023}. Thus, the experimentally observed~\cite{crsbr:exp} large and nearly isotropic diffusion of excitons in CrSBr can be understood in terms of the magnon-exciton drag effect delineated here. Our treatment contrasts with previous approaches to magnetic excitons propagation~\cite{eremenko2012magneto,Yamamoto:1977aa,PhysRevLett.41.1681,Ueda:1980aa} that were mainly based on the hopping effects and highlights the specifics of van der Waals systems like CrSBr.

Finally, in Sec.~\ref{sec:discussion}, we discuss some limitations of our analysis in this work and potential ways to address them in the future. In particular, a negative magnon-exciton drag due to the negatively dispersing dipolar magnons and superdiffusive exciton transport demonstrated in Ref.~\cite{crsbr:exp} under far-from-equilibrium conditions are not captured by the quasiequilibrium analysis presented here. We close with concluding remarks in Sec.~\ref{sec:conclusion}.

\section{Magnons in {C\MakeLowercase{r}SB\MakeLowercase{r}} films}\label{sec:magnon}

	



\subsection{General formalism}

We briefly outline the calculation of the magnon dispersion following Ref.~\cite{crsbr:exp}. Let us start with a bilayer structure. The most general approach is to use the lattice Hamiltonian of Cr ions in the form
\begin{equation}
    \label{H:magnetic:tot}
    \mathcal H_{M} = \mathcal H^{FM}+ \mathcal H^{AFM} + \mathcal H^a + \mathcal H^{dd},
\end{equation}
where $\mathcal{H}^{FM}$ describes the ferromagnetic interactions of the spins in a given layer, $\mathcal H^{AFM}$ describes the weak antiferromagnetic coupling of the layers, $\mathcal H^a$ describes single-ion anisotropies, and $\mathcal H^{dd}$ is the dipole-dipole interaction between the lattice spins. We neglect for a moment the dipole-dipole coupling. The ferromagnetic interaction Hamiltonian reads
\begin{equation}
    \mathcal{H}^{FM} = \sum_{i,j}J_{\langle ij\rangle}\left(\bm S^{(1)}_i\bm S^{(1)}_j + \bm S^{(2)}_i\bm S^{(2)}_j\right),
    \label{FMham}
\end{equation}
where $\bm S^{(1)}_i$ and $\bm S^{(2)}_j$ are the spin operators at the sites (Cr ions) $i$ and $j$, superscripts $(1)$ and $(2)$ denote the layers, $J_{\langle ij\rangle}$ are the exchange constants. The interlayer antiferromagnetic exchange interaction is taken into account only for the sites one on top of the other,
\begin{equation}
    \mathcal{H}^{AFM} = \sum_iJ_{int}\bm S^{(1)}_i\bm S^{(2)}_i,
\end{equation}
with a single parameter $J_{int}$\footnote{The stacking of CrSBr is of A-type where the Cr atoms are one on top of the other.}, and the crystal anisotropy with the easy $y$-axis can be presented as
\begin{multline}  
    \mathcal{H}^{a} = \sum_iK_x\left(S^{(1)}_{i,x}S^{(1)}_{i,x} + S^{(2)}_{i,x}S^{(2)}_{i,x}\right) \\+ \sum_iK_z\left(S^{(1)}_{i,z}S^{(1)}_{i,z} + S^{(2)}_{i,z}S^{(2)}_{i,z}\right).
\end{multline}

For the following it is instructive to introduce the Fourier transforms of the interactions. Accordingly, the Fourier transform of the ferromagnetic interaction reads (the sign is selected for convenience)
    \begin{equation}
        J^{FM}(\bm k) \equiv -\sum_{j-i}{J}_{\langle j-i\rangle}e^{{\rm i}\bm k\bm r_{j-i}},
    \end{equation}
 where $\bm r_{j-i} \equiv \bm r_j - \bm r_i$, and $\bm r_i, \bm r_j$ are the in-plane coordinates of the ions. The antiferromagnetic interaction is momentum-independent and is also isotropic in the spin space.

Note that  the dominant $J_{\langle ij\rangle} < 0$, while $J_{int} > 0$, $K_x > 0$ and $K_z > 0$. In our model in the equilibrium the spins in the first and second layers are saturated along and opposite the $y$-axis. The analysis of the magnetic field effects on the magnon dispersion and transport are beyond the scope of this work.

We perform the Holstein-Primakoff transformation 
\begin{subequations}
\label{HS:AF}
\begin{align}
&S_-^{(1)} = \sqrt{2S} a^\dag, \quad S_+^{(1)} = \sqrt{2S} a, \quad S_y^{(1)} = S - a^\dag a, \nonumber \\ 
&S_x^{(1)} = \frac{1}{\mathrm i} \sqrt{\frac{S}{2}}(a - a^\dag), \quad S_z^{(1)} =  \sqrt{\frac{S}{2}}(a + a^\dag),\\
&S_-^{(2)} = \sqrt{2S} b, \quad S_+^{(2)} = \sqrt{2S} b^\dag, \quad S_y^{(2)} = -S + b^\dag b, \nonumber \\  
&S_x^{(2)} = \frac{1}{\mathrm i} \sqrt{\frac{S}{2}}(b^\dag - b), \quad S_z^{(2)} =  \sqrt{\frac{S}{2}}(b + b^\dag),
\end{align}
\end{subequations}
where the subscripts enumerating the lattice sites are omitted for brevity to introduce the operators $a^\dag,a,b^\dag,b$ which obey the Bose commutation rules. This transformation is approximate and holds for not too high magnon occupancies, where higher-order in $a,a^\dag,\ldots$ terms can be omitted. Making the Fourier transform of the resulting Hamiltonian as, e.g.,
\begin{equation}
    a_{\bm k} = \sum_j\frac{e^{-{\rm i}\bm k\bm r_j}}{\sqrt{\mathcal{N}}}a_j, \quad a_j = \sum_{\bm k}\frac{e^{{\rm i}\bm k\bm r_j}}{\sqrt{\mathcal{N}}}a_{\bm k},
\end{equation}
where $\mathcal{N}$ is the number of Cr ions in the layer plane (twice the number of unit cells in the plane),
we arrive at the energies of two magnon branches~\cite{crsbr:exp}
\begin{equation}
\label{energ:gen}
    E^\pm_{\bm k} = \frac1{2}\sqrt{(B_{\bm k} \pm D_{\bm k})^2 - (A_{\bm k} \pm C_{\bm k})^2},
\end{equation}
where the coefficients read
\begin{subequations}
\label{AD}
    \begin{align}
        A_{\bm k} & = 2SJ_{int}, \\
        B_{\bm k} & = 2S\left[J_{int}+K_x+K_z+J^{FM}(0) - J^{FM}(\bm k)\right], \\
        C_{\bm k} & = 2S(K_z - K_x), \\
        D_{\bm k} & = 0.
    \end{align}
\end{subequations}

Note that the Hamiltonian of magnetic system is diagonal in the basis of the operators $e,e^\dag,f,f^\dag$ which are linear combinations of the operators $a,a^\dag,b,b^\dag$:
\begin{equation}
    \label{H:magn:diag}
    \mathcal H_M = \sum_{\bm k} (E_{\bm k}^+ e^\dag_{\bm k} e_{\bm k} + E_{\bm k}^- f^\dag_{\bm k} f_{\bm k}),
\end{equation}
where we excluded the ground state energy from $\mathcal H_M$ and
\begin{equation}
\label{transition:fin}
        \begin{pmatrix}
        a_{\bm k} \\ b_{\bm k} \\ a^\dag_{-\bm k} \\ b^\dag_{-\bm k}
    \end{pmatrix} = \frac1{2}
    \begin{pmatrix}
        X_{\bm k}^{++} & -X_{\bm k}^{-+} & -X_{\bm k}^{+-} & -X_{\bm k}^{--} \\
        X_{\bm k}^{++} & X_{\bm k}^{-+} & -X_{\bm k}^{+-} & X_{\bm k}^{--} \\
        -X_{\bm k}^{+-} & -X_{\bm k}^{--} & X_{\bm k}^{++} & -X_{\bm k}^{-+} \\
        -X_{\bm k}^{+-} & X_{\bm k}^{--} & X_{\bm k}^{++} & X_{\bm k}^{-+}
    \end{pmatrix}\begin{pmatrix}
        e_{\bm k} \\ f_{\bm k} \\ e^\dag_{-\bm k} \\ f^\dag_{-\bm k}
    \end{pmatrix}
\end{equation}
with the coefficients
\begin{equation}
\label{X:coeff}
    X_{\bm k}^{\pm\pm'} = \sqrt{\frac{B_{\bm k} \pm D_{\bm k}}{2E_{\bm k}^\pm} \pm' 1}.
\end{equation}
Note that for the actual parameters of CrSBr the $X^{--}_{\bm k} < 0$, hence, for this particular coefficient the square root should be taken with the negative sign.
While Eqs.~\eqref{energ:gen}, \eqref{H:magn:diag} and \eqref{transition:fin} were derived neglecting the dipole-dipole interactions, these expressions can be used in the general case as well with appropriately renormalized parameters as shown below. \addZakhar{The transformation coefficients are analyzed in detail in Appendix~\ref{app:Xs}.}

\subsection{Dipole-dipole interaction}\label{sec:dd}

We now take into account the long-range dipole-dipole interaction between the Cr spins. These effects are particularly pronounced for small wavevectors compared to the Brillouin zone dimensions. Here we consider a bilayer case based on a discrete model, the dipole-dipole interaction effects in multilayers are analyzed within a macroscopic approach in Appendix~\ref{app:semiclass}. Physically, the dipole-dipole interaction arises as a result of the spin interaction with the magnetic field produced by the net magnetization in the system. Since the interlayer distance in the bilayer is small compared to the wavelength of relevant magnons (and, naturally, with the wavelength of the produced electromagnetic field), it is sufficient to consider only the excitations for which the effective ``total'' spin of the layers is nonzero, 
\begin{equation}
    \bm S^{eff}_i = \bm S^{(1)}_i + \bm S^{(2)}_i \ne 0.
\end{equation}
Here the subscript $i$ enumerates the pairs of Cr ions one on top of the other.
The dipole-dipole interaction Hamiltonian has the general form
\begin{equation}
    \label{Hdd_discrete}
    \mathcal{H}^{dd} = -\frac{\hbar^2\gamma^2}{2}\sum_{i\neq j}\frac{\Xi_{ij}}{|\bm r_j - \bm r_i|^5},
\end{equation}
with $\gamma = 2\mu_B / \hbar (s / S)$ (where $s \approx 3.56/2$~\cite{Scheie:2022aa}) being the effective gyromagnetic ratio for both layers together, and
\[
\Xi_{ij} = 3\left(\bm S^{eff}_i\cdot\bm r_{j-i}\right)\left(\bm S^{eff}_j\cdot \bm r_{j-i}\right) - \bm S^{eff}_i\cdot \bm S^{eff}_j|\bm r_{j-i}|^2.
\]
By virtue of the linearized Holstein-Primakoff transformation the effective spin can be expressed as
\begin{align}
    \label{Seff}
    S^{eff}_x = \frac1{\rm i}\sqrt{\frac{S}{2}}\left(a - b - a^\dag + b^\dag\right), \quad S^{eff}_y = 0, \\
    S^{eff}_z = \sqrt{\frac{S}{2}}\left(a + b + a^\dag + b^\dag\right),\nonumber
\end{align}
where, as above, the lattice site subscripts were omitted. Note that quadratic in $a, a^\dag$, $b,b^\dag$ contributions to $S_y^{eff}$ can be omitted for the analysis of the dipole-dipole interaction.
Following Ref.~\cite{crsbr:exp} we obtain the additive contributions of the dipole-dipole interactions to the parameters $A_{\bm k} \ldots D_{\bm k}$
\begin{subequations}
\label{AD:dd}
    \begin{align}
        A^{dd}_{\bm k} & = -2S\hbar u\frac{k^2 - k_x^2}{k}, \\
        B^{dd}_{\bm k} & = -2S\hbar u\frac{k^2 - k_x^2}{k}, \\
        C^{dd}_{\bm k} & = -2S\hbar u\frac{k^2 + k_x^2}{k}, \\
        D^{dd}_{\bm k} & = -2S\hbar u\frac{k^2 + k_x^2}{k}.
    \end{align}
\end{subequations}
where we introduced the constant of the dimensionality of velocity
\begin{equation}
\label{velocity}
    u = {2}\frac{\pi \hbar \gamma^2}{\mathcal{A}_0},
\end{equation}
and the unit cell area $\mathcal A_0$ (note that there are 4 Cr atoms in total in the unit cell of bilayer). The energies of the magnons are given by Eq.~\eqref{energ:gen} with the constants being the sum of expressions \eqref{AD} and \eqref{AD:dd}. The transformation coefficients are given by~\eqref{transition:fin}.
The branch $E^+_{\bm k}$ is polarized along the $z$-axis, i.e. its effective magnetization $\propto \bm S^{eff}$ is parallel to $z$-axis, and the branch $E^-_{\bm k}$ corresponds to $\bm S^{eff} \parallel x$. The dispersion of magnons can also be derived from the semiclassical equations of motion of  magnetizations, see Appendix~\ref{app:semiclass}. Note that Eqs.~\eqref{AD:dd} and the analysis presented here holds for all wavevectors apart from a very small range of $k<E^\pm_0/(\hbar c)$ where retardation effects may be important. Our estimates show that the retardation is irrelevant for the following.

\begin{table*}[ht]
    \centering
        \caption{Set of parameters used in calculation, see corresponding sections in the text for details, see Refs.~\cite{Bo_2023,Scheie:2022aa,Bae:2022,Semina:2025}.
        }
    \label{tab:param}
    {
    \begin{tabular}{|c|c|c|c|c|c|c|}
        \hline
        \rowcolor{lightgray!50}
        $a$ &
        $b$ &
        $S$ &
        $\hbar\gamma$ &
        $J_{int}$ &
        $K_x$ &
        $K_z$ \\ \hline
        
        $3.5$~\r{A}
        & $4.76$~\r{A}
        & $3/2$
        & $2\mu_B \frac{3.56}{2S}$
        & $6~\mu$eV
        & $14~\mu$eV
        & $58~\mu$eV \\ \hline\hline
        
        \rowcolor{lightgray!50}
        $J_1$ &
        $J_2$ &
        $J_3$ &
        $J_4$ &
        $J_5$ &
        $J_6$ &
        $J_7$ \\ \hline

        $-1.9$~meV
        & $-3.38$~meV
        & $-1.67$~meV
        & $-0.09$~meV
        & $-0.09$~meV
        & $0.37$~meV
        & $-0.29$~meV \\ \hline\hline
        
        \rowcolor{lightgray!50}
        $m_m^x$ &
        $m_m^y$ &
        $E_0^M$ &
        $m_e^x$ &
        $m_e^y$ &
        $m_h^x$ &
        $m_h^y$ \\ \hline

        $73m_0$
        & $23m_0$
        & $0.1$~meV
        & $7.31m_0$
        & $0.14m_0$
        & $2.84m_0$
        & $0.45m_0$ \\ \hline\hline
        
        \rowcolor{lightgray!50} 
        $M_x$ &
        $M_y$ &
        $E_D$ &
        $E_I$ &
        $|t_e|$ &
        $r_x$ &
        $r_y$ \\ \hline

        $10.15m_0$
        & $0.59m_0$
        & $200$~meV
        & $80$~meV
        & $53$~meV
        & $0.6$~nm
        & $1.8$~nm \\ \hline
    \end{tabular}}
\end{table*}


\begin{figure}
    \centering
    \includegraphics[width=\linewidth]{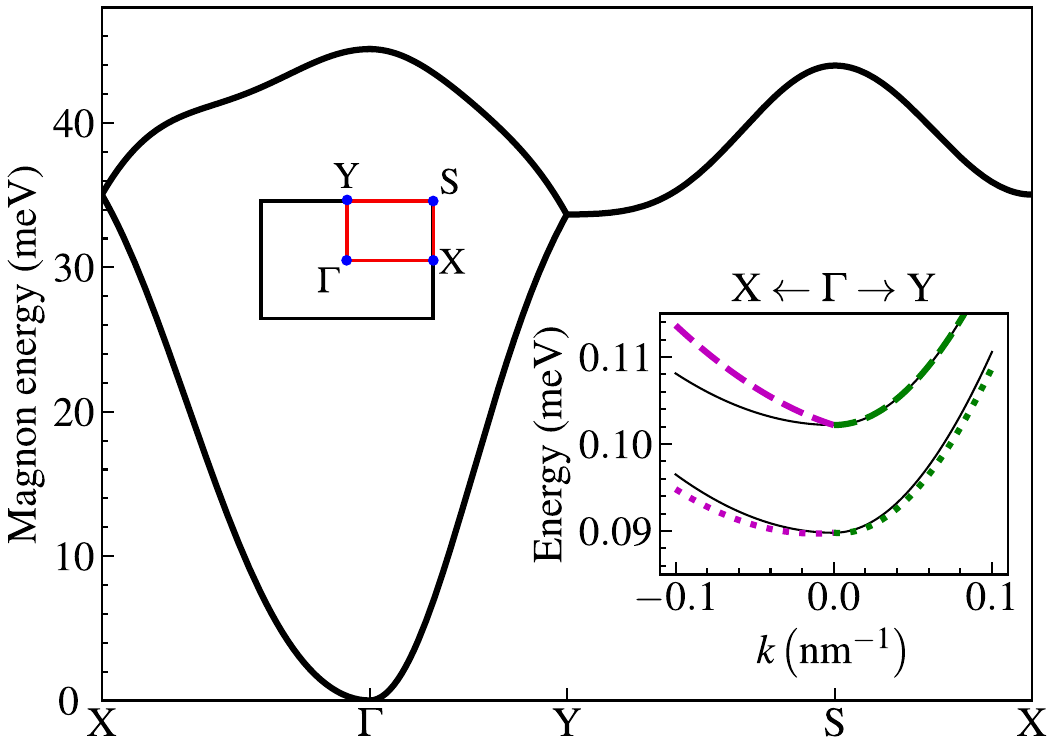}
    \caption{Magnon dispersion for the full Brillouin zone. On this scale two magnon modes are indistinguishable. Inset shows magnon dispersion near the $\Gamma$-point, left magenta lines correspond to the $\Gamma\to X$ direction, right green lines correspond to $\Gamma \to Y$ direction. Dotted and dashed lines show $E^+_{\bm k}$ and $E^-_{\bm k}$, respectively, Eq.~\eqref{energ:gen}. Thin solid black lines show dispersion without dipole-dipole interaction. Inset shows CrSBr Brillouin zone.}
    \label{fig:disp}
\end{figure}

Magnon dispersion calculated across the whole Brillouin zone using Eqs.~\eqref{energ:gen} with allowance for the dipole-dipole interaction is shown in Fig.~\ref{fig:disp}. The parameters of intralayer interaction $J_{\langle ij\rangle}$, Eq.~\eqref{FMham}, are taken from Ref.~\cite{Scheie:2022aa}: interlayer interaction $J_{int} = 6~\mu$eV and single-ion anisotropy parameters $K_x = 14~\mu$eV and $K_z = 58~\mu$eV are taken from Ref.~\cite{Bae:2022} (the complete set of used parameters is summarized in Tab.~\ref{tab:param}). Overall, the dispersion is in good agreement with previous works~\cite{Scheie:2022aa,Bae:2022,w4xn-2yff} with minor differences resulting from the variation of parameters. In this scale the splitting between the branches $E^+_{\bm k}$ and $E^-_{\bm k}$ is indistinguishable. The inset demonstrates the impact of dipole-dipole interaction in bilayer CrSBr and the splitting between the branches in the vicinity of the $\Gamma$-point. As a result of the dipole-dipole interaction, the bottom magnon energy branch ($E^+_{\bm k}$, $z$-polarized) for very small wavevectors has an isotropic dispersion $\propto u_+ k$ with the negative velocity
\begin{equation}
\label{u+}
u_+ = - 2S \sqrt{\frac{K_x}{K_z+J_{int}}} u,
\end{equation}
while the top magnon branch ($x$-polarized) is anisotropic with $\Delta E^-_{\bm k} \propto k_x^2 / k$. The effect is quite weak in bilayer system. The effect of the dipole-dipole interaction increases with the increase in the sample thickness. As demonstrated in Appendix~\ref{app:FiniteThickness} in the sample consisting of $N$ CrSBr bilayers ($N=L/d$ where $d$ is the effective bilayer thickness and $L$ is the sample thickness) the velocity of the bottom, $z$-polarized magnon branch $u_+(N) \approx N u_+$, where $u_+$ is introduced in Eq.~\eqref{u+}. The linear dispersion of the bottom branch is valid for $kL \ll 1$.

\section{Exciton-Magnon coupling}\label{sec:coupling}

In this section, at first, we qualitatively describe the mechanism of electron-magnon and exciton-magnon interaction in quasi-classical way (Sec.~\ref{subsec:quasicl}) and then we build a microscopic theory in Sec.~\ref{subsec:quantum}. The interaction has an orbital nature. In equilibrium, the magnetizations of the neighboring layers are antiparallel. The electrons and holes are unable to tunnel between the layers. The magnons, similarly to the external magnetic field, tilt the magnetization of the layers and allow the charge carriers tunneling (Fig.~\ref{fig:mixing}). It results in the mixing of electronic states in the layers and, consequently, in the variation of the exciton energy.

\subsection{Quasi-classical approach}\label{subsec:quasicl}

\begin{figure}[ht]
    \centering
    \includegraphics[width=\linewidth]{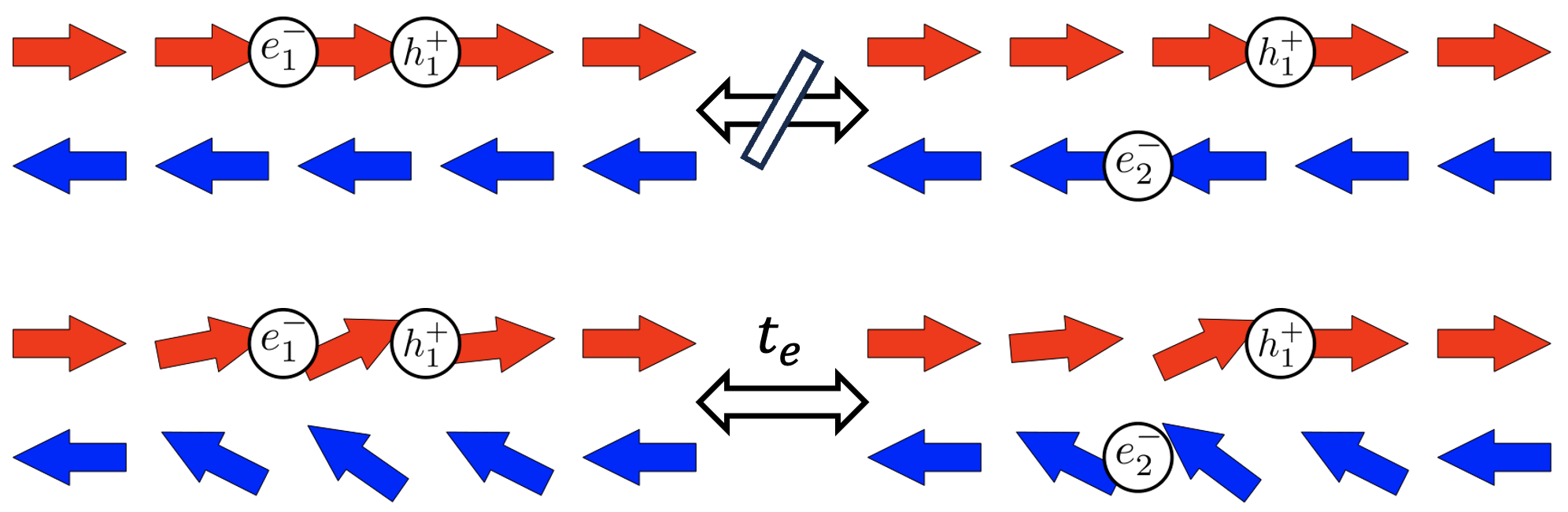}
    \caption{In equilibrium, the magnetization of the neighboring layers is antiparallel and the states of electrons are orthogonal. In the presence of magnons the magnetization is tilted and the mixing of intralayer and interlayer exciton is allowed.}
    \label{fig:mixing}
\end{figure}

In the quasi-classical approach we consider the magnetizations of the layers as classical vectors which smoothly depend on the coordinate and assume that the charge carriers spins follow local direction of magnetization. We start with the case of a single charge carrier electron, $e$, or hole, $h$. The tunneling matrix element is
\begin{equation}
    \label{eq:Vqc}
    \mathcal{V}_{e,h} = {t_{e,h}} \left\langle\left. S_1^{e,h}\right|S_2^{e,h}\right\rangle = {t_{e,h}} \cos{\left(\frac{\theta(\bm r)}{2}\right)}.
\end{equation}
Here $|S_i^{e,h}\rangle$ ($i=1,2$) are the spin states of the charge carrier in a given layer, $\theta(\bm r)$ ($|\theta- \pi| \ll 1$) is the angle between the layer magnetizations, and $t_{e,h}$ is the tunneling constant related to an overlap of the orbital wavefunctions in the neighboring layers. For simplicity we assume that $|{t}_e| \gg |{t}_h|$ and focus on the electron tunneling only. 
The interaction~\eqref{eq:Vqc} mixes the intra- and interlayer excitons (see Refs.~\cite{Tabataba-Vakili:2024aa,Semina:2025} for further details on excitonic structure), see Fig.~\ref{fig:mixing}. As a result of the mixing, the intralayer exciton acquires a coordinate-dependent energy shift 
\begin{equation}
\label{shift:class}
    \Delta E(\bm r) = -\frac{{t}^2_{e}}{E_D - E_I}\cos^2{\left(\frac{\theta(\bm r)}{2}\right)},
\end{equation}
where $E_D, E_I > 0$ are the binding energies of intralayer and interlayer excitons. The constants can be estimated from experimental result~\cite{Wilson:2021aa}, where the angle $\theta$ was controlled by an external magnetic field, and numerical calculations of the exciton binding energies~\cite{Semina:2025} as $E_D = 200$~meV, $E_I = 80$~meV and ${t}_{e} = 53$~meV. 
Equation~\eqref{shift:class} shows that the deviation of the magnetization from the equilibrium antiferromagnetic configuration, $\theta \ne \pi$, results in the variation of the exciton energy, i.e., in the exciton-magnon interaction.
Qualitatively, in the absence of external magnetic field, the matrix element of the mixing~\eqref{eq:Vqc} $\propto \cos{[\theta(\bm r)/2]}$ and on the quantum language this process is single-magnon. At the same time, the magnon interaction with the intralayer exciton appears in the second order of perturbation and has a two-magnon nature, as we demonstrate rigorously below in quantum mechanical approach.

\subsection{Quantum mechanical approach}\label{subsec:quantum}

Let us now derive the quantum mechanical expression for the exciton-magnon interaction in the second quantization approach. To that end we consider the exchange interaction of the electron with the Cr spins in the isotropic form of the electron and Cr spin operators
\begin{multline}
    \mathcal H_{e-Cr} = - \sum_{i}\frac{\mathcal{A}_0E_S}{2}\delta(\bm r - \bm r_i)\,\hat{\bm s}\cdot \left(\bm S_i^{(1)}+\bm S_i^{(2)}\right) \\ = \sum_{i}\frac{\mathcal{A}_0}{2}\delta(\bm r - \bm r_i)\left(\mathcal H_{S,i}^{(1)}+ \mathcal H_{S,i}^{(2)}\right),
\end{multline}
where $E_S>0$ is the parameter related to the exchange-induced splitting of the conduction band, $\hat{\bm s}$ is the electron spin operator. Note that there are 2 Cr ions in the unit cell per layer, $\bm r$ is the electron coordinate and $i$, as before, enumerates the Cr ions in one layer. By means of Eqs.~\eqref{HS:AF} we obtain\footnote{We used here the representation of the spin operators such that $\sigma_z$ Pauli matrix corresponds to $2\hat{s}_y$ (the other components are obtained by cyclic permutations of Cartesian subscripts) to simplify the form of resulting expressions.}
\begin{subequations}
    \label{eq:Hs}
\begin{equation}
    \mathcal{H}^{(1)}_{S,i} 
    = -\frac{E_S}{2}\begin{pmatrix}
        S-a^\dagger_ia_i & \sqrt{2S}a^\dagger_i \\
        \sqrt{2S}a_i & -S + a^\dagger_ia_i
    \end{pmatrix},
\end{equation}
\begin{equation}
    \mathcal{H}^{(2)}_{S,i} 
    = -\frac{E_S}{2}\begin{pmatrix}
        -S+b^\dagger_ib_i & \sqrt{2S}b_i \\
        \sqrt{2S}b^\dagger_i & S - b^\dagger_ib_i
    \end{pmatrix}.
\end{equation}
\end{subequations}

Now we need to take into account the interlayer tunneling. As above, the tunneling is characterized by the tunneling constant $t_e$, the interlayer tunneling conserves the spin and wavevector of the electron. It can be chosen real. Introducing the four-component electron wavefunction $\Psi = \left(\Psi^{(1)}_{\uparrow},\Psi^{(1)}_{\downarrow},\Psi^{(2)}_{\uparrow},\Psi^{(2)}_{\downarrow}\right)^T$, where arrows $\uparrow$ and $\downarrow$ correspond to the electron spin oriented along the $y$-axis and opposite to it and superscripts $1$ and $2$ denote the layer where the electron is localized, we obtain the effective Hamiltonian of the electron-Cr interaction in the form
\begin{equation}
    \mathcal{H}_{e,i} = E_S\begin{pmatrix}
        -\frac{S-a^\dagger_ia_i}{2} & -\sqrt{\frac{S}{2}}a^\dagger_i & \frac{t_e}{E_S} & 0 \\
        -\sqrt{\frac{S}{2}}a_i & \frac{S - a^\dagger_ia_i}{2} & 0 & \frac{t_e}{E_S} \\
        \frac{t_e}{E_S} & 0 & \frac{S-b^\dagger_ib_i}{2} & -\sqrt{\frac{S}{2}}b_i \\
        0 & \frac{t_e}{E_S} & -\sqrt{\frac{S}{2}}b^\dagger_i & -\frac{S - b^\dagger_ib_i}{2}
    \end{pmatrix}.
\end{equation}

We can now integrate out two electron states $\Psi^{(1)}_{\downarrow}$ and $\Psi^{(2)}_{\uparrow}$ which are much higher in energy compared to the pair of states $\Psi^{(1)}_{\uparrow}$ and $\Psi^{(2)}_{\downarrow}$: For the latter, the spins are parallel to the  magnetizations of the corresponding layers, while for the former the spins are antiparallel to the corresponding magnetizations\addZakhar{, see Appendix~\ref{app:technical:x-m} for details.}
Keeping only the linear in the magnon operators contributions and choosing the origin of the energy as the energy of the unperturbed system we arrive at the Hamiltonian
\begin{equation}
\label{H:e:int}
    \mathcal{H}_{e,i} = \frac{t_e}{\sqrt{2S}}\begin{pmatrix}
        0 & c_i^\dagger \\
        c_i & 0
    \end{pmatrix}
\end{equation}
acting in the ground states subspace $\Psi = \left(\Psi^{(1)}_{\uparrow},\Psi^{(2)}_{\downarrow}\right)^T$. To shorten the notations we introduced the operators,
\begin{equation}
    c_i = b_i^\dagger + a_i,  \qquad c_i^\dagger = b_i + a_i^\dagger,
\end{equation}
that obey the commutation relation $[c_i^\dagger, c_j] = 0$.

Turning to the exciton-magnon interaction, we recall that there are two types of exciton states in the bilayer: intralayer excitons and interlayer excitons. Accordingly, the exciton Hamiltonian that accounts for the electron interaction with Cr ions (Fig.~\ref{fig:mixing}) takes form
\begin{equation}
    \mathcal{H}_{X-Cr} = -\begin{pmatrix}
        E_D & 0 \\
        0 & E_I
    \end{pmatrix} + \sum_i\frac{\mathcal{A}_0}{2}\delta(\bm r - \bm r_i)\mathcal{H}_{e,i}.
\end{equation}
Here, $E_D$ and $E_I$ are, respectively, the binding energies of the intra- and interlayer excitons (note the minus sign in front of the Hamiltonian). Here we keep only the  exciton states with the hole in one particular layer.

In antiferromagnetic state, the interlayer and intralayer excitons are well separated. In this work we are mainly interested in the dynamics of intralayer excitons, that are observed in experiments. We introduce the splitting
\begin{equation}
    \Delta = E_D - E_I>0,
\end{equation}
between the inter- and intralayer excitons and take into account that $\Delta$ is much greater than kinetic energies and tunneling constants. Thus, we exclude the interlayer excitons in the second order perturbation theory similar to Eq.~\eqref{eq:H41} \addZakhar{of Appendix~\ref{app:technical:x-m}} with the resulting effective Hamiltonian which now, naturally, involves two Cr ions $i$ and $j$
\begin{equation}
    \label{HXM-real}
    \mathcal{H}_{XM,D} = -\sum_{i,j}\frac{t_e^2}{2S\Delta}c_j^\dagger c_i \hat{H}_{D,ij},
\end{equation}
where the operator $\hat{H}_{D,ij}$ acts on the intralayer exciton envelope functions and takes the form
\begin{equation}
    \label{HDij}
    \hat{H}_{D,ij} = \left(\frac{\mathcal{A}_0}{2}\right)^2\sum_I\delta(\bm r_{e,1} - \bm r_j) \left|\Psi_I\right\rangle \langle\Psi_I|\, \delta(\bm r_{e,2} - \bm r_i),
\end{equation}
and the summation over all intermediate states of interlayer exciton $I$ is assumed, $\left|\Psi_I\right\rangle$ and $\langle\Psi_I|$ and the ket- and bra-states of the interlayer excitons, $\bm r_{e,1}$ and $\bm r_{e,2}$ are the electron coordinates related to the ket- and bra-states, respectively.  For convenience we have also excluded the diagonal part $E_D$ which is independent of the magnon operators.

Finally, we need to rewrite the Hamiltonian in momentum space. We use exciton envelope wavefunctions in the form
\begin{equation}
    \Psi_{D, \bm k} (\bm R, \bm \rho) = \frac{e^{{\rm i}\bm k\bm R}}{\sqrt{\mathcal S}}\varphi_D(\bm \rho), \qquad \Psi_{I, \bm k} (\bm R, \bm \rho)  = \frac{e^{{\rm i}\bm k\bm R}}{\sqrt{\mathcal S}}\varphi_I(\bm \rho),
\end{equation}
where $\mathcal S$ is the normalization area $\bm R =(X,Y)$ and $\bm \rho=(x,y) = \bm r_e - \bm r_h$ are the center of mass and relative motion in-plane coordinates of the electron-hole pair, $\varphi_{D,I}(\bm \rho)$ are the relative motion envelope functions. Due to the anisotropy of electronic bands of CrSBr the effective masses of electron, $m_e^\alpha$, hole, $m_h^\alpha$, and exciton $M^\alpha = m_e^\alpha+m_h^\alpha$ are different for the main axes $\alpha = x$ and $y$ of the layer, accordingly, $X=(m_e^x x_e + m_h^x x_h)/M^x$ and $Y=(m_e^y y_e + m_h^y y_h)/M^y$. We also introduce the form-factor to correctly take into account the tunneling to the different orbital states
\begin{equation}
    \label{Feq}
    \mathcal{F}^e_{\bm q} = \int d\bm \rho \varphi_I^*(\bm \rho)\varphi_D(\bm \rho)e^{-{\rm i}\bm q\bm\rho m_h/M},
\end{equation}
where the expression
\[
\frac{\bm q\bm \rho m_h}{M}\equiv \sum_{\alpha=x,y} q_\alpha \rho_\alpha \frac{m^\alpha_h}{M^\alpha},
\]
and effective tunneling parameters are introduced as
\begin{equation}
   t^e_{\bm q} = t_e\mathcal{F}^e_{\bm q}.
\end{equation}
These parameters can be chosen real because ground state envelope functions are real even functions of $\bm \rho$. We take into account the intermediate states with the ground-state envelope assuming that the remaining states are much higher in the energy. As a result we obtain the following Hamiltonian describing the scattering of the exciton with the wavevector $\bm k_1$ to the state with the wavevector $\bm k_2$ read
\begin{equation}
\label{HXM-momentum1}
\mathcal{H}_{XM}(\bm k_2, \bm k_1) = -{\frac{1}{\mathcal{N}}}\sum_{\bm k'}{c^\dagger_{\bm k' - \bm k_2}}\frac{t_{\bm k_2 - \bm k'}^{e*} t^e_{\bm k' - \bm k_1}}{2S\Delta}{c_{\bm k' - \bm k_1}},
\end{equation}
where, as above,
\begin{equation}
\label{c:q:def}
c_{\bm q} = b^\dagger_{-\bm q} + a_{\bm q}, \qquad c^\dagger_{\bm q} = b_{-\bm q} + a^\dagger_{\bm q}.
\end{equation}

The analysis of the hole tunneling can be carried out in exactly the same way and results in the interaction Hamiltonian in the same form as~\eqref{H:e:int} with the replacement of $t_e$ by $t_h$, the hole tunneling constant. As a result, we arrive at the following $2\times 2$ Hamiltonian acting in the basis of two intralayer excitons in the first and second layers, respectively:
\begin{equation}
\label{HXM-momentum2}
\mathcal{H}_{XM}(\bm k_2, \bm k_1)  = \begin{pmatrix}
  \mathcal{H}_{XM}(\bm k_2, \bm k_1)_{11} & \mathcal{H}_{XM}(\bm k_2, \bm k_1)_{12} \\
   \mathcal{H}_{XM}(\bm k_2, \bm k_1)_{21} & \mathcal{H}_{XM}(\bm k_2, \bm k_1)_{22}
\end{pmatrix},
\end{equation}
where the matrix elements \addZakhar{are presented in Eq.~\eqref{HXM-momentum2:me} of Appendix~\ref{app:technical:x-m}}.

The physical meaning of the Hamiltonian is simple: the magnon tilts the magnetization of the layers and enables mixing of the intra- and interlayer excitons via the charge carrier tunneling, Fig.~\ref{fig:mixing}. In the second-order perturbation theory it produces the energy variations of the excitons. To demonstrate all relevant effects, it is sufficient for only one type of carriers to tunnel, and we focus here on this scenario. In this case, there is no mixing between the intralayer excitons in the first and second layers, correspondingly, $\mathcal{H}_{XM}(\bm k_2, \bm k_2)_{12}$ and $\mathcal{H}_{XM}(\bm k_2, \bm k_2)_{21}$, vanish as they requires both nonzero tunneling constants $t_e$ and $t_h$\addZakhar{, see Eq.~\eqref{HXM-momentum2:me}}. This is because such mixing requires co-tunneling of an electron and a hole. The matrix elements of exciton-magnon interaction of intralayer excitons are non-zero even if only electron or hole can tunnel between the layers, they differ by numerical values and the phases of magnon operators because of the symmetry between the layers. In what follows we consider for specificity that only one type of carriers, electrons, can tunnel between the layers. 

We emphasize, that all interaction processes described by the Hamiltonian~\eqref{HXM-momentum2} have two-magnon nature in agreement with qualitative picture presented above. In the presence of external magnetic field the antiferromagnetic configuration is already broken in the equilibrium and one-magnon processes are important. There is an analogy between exciton-magnon and exciton-flexural phonon interaction in van der Waals monolayers and bilayers. Similar to the magnetic case, the electron- and exciton-flexural phonon is suppressed in monolayer structures due to the symmetry consideration, as charged carriers move in plane, while the flexural phonon-induced monolayer displacement is orthogonal to it. The situation is opposite for the interlayer excitons in bilayers: the breathing modes of flexural phonons modify the interparticle distance, giving rise to the effective interaction with the change in the Coulomb binding energy~\cite{Semina2020Dec,Iakovlev2022}.

\section{Exciton-magnon polaron}\label{sec:polaron}

The first consequence of the exciton-magnon interaction is the dressing of excitons with magnons, i.e., formation of the exciton-magnetic or exciton-magnon polaron. This effect is somewhat similar to the formation of magnetic polaron in conventional magnetic semiconductors where the exchange interaction between the charge carrier spin and the magnetic subsystems results in the reduction of the energy and autolocalization of the carrier~\cite{PhysRev.118.141,nagaev:polaron_eng,PhysRevB.31.8024,Wolff:1988aa,Nagaev:1992aa,merkulov:polaron,PhysRevB.51.14124}. Here we present the analysis of magnetic polaron based on the interaction above within the perturbation theory. We demonstrate that the effect is present but does not lead to strong modifications of excitonic parameters, making it possible to neglect it in the analysis of the magnon-exciton drag.

Within the framework of the perturbation theory the leading order polaron effect can be evaluated by the diagonal components of Hamiltonian~\eqref{HXM-momentum2} $\mathcal{H}_{XM}(\bm k, \bm k)$ for the fixed magnon population. We consider equilibrium situation at temperature $T$ where the following relations hold:
\begin{subequations}
    \label{thermal:magnon}
\begin{align}
    \langle e^\dag_{\bm q} e_{\bm q} \rangle = \langle e_{\bm q} e_{\bm q}^\dag \rangle - 1 = \frac{1}{\exp{\left(\frac{E_{\bm q}^+}{k_B T} \right)}-1} \equiv n^e_{\bm q},\\
    \langle f^\dag_{\bm q} f_{\bm q} \rangle = \langle f_{\bm q} f_{\bm q}^\dag \rangle - 1 = \frac{1}{\exp{\left(\frac{E_{\bm q}^-}{k_B T} \right)}-1}  \equiv n^f_{\bm q},
\end{align}
\end{subequations}
where $k_B$ is the Boltzmann constant and all remaining second-order averages are zero. Using Eqs.~\eqref{transition:fin} and \eqref{c:q:def} we obtain
\begin{equation}
    \label{EXMP}
    {\left\langle \mathcal{H}_{XM}\right\rangle = -{\frac{1}{\mathcal{N}}}\sum_{\bm q}\mathcal N_{\bm q}
        \frac{\left|t^e_{\bm q}\right|^2}{4S\Delta},}
\end{equation}
where
\begin{multline}
\label{magnon:N:q}
    \mathcal N_{\bm q} = (X^{++}_{\bm q} - X^{+-}_{\bm q})^2\left(n^e_{\bm q} + \frac{1}{2}\right)  \\ + (X^{-+}_{\bm q} - X^{--}_{\bm q})^2\left(n^f_{\bm q} + \frac{1}{2}\right).
\end{multline}
The quantity $\mathcal N_{\bm q}$ provides effective magnon population with account for the coefficients $X_{\bm q}^{\pm\pm}$ that allow for the specifics of magnon-induced magnetization variation in our system.
Although Eq.~\eqref{EXMP} was derived neglecting exciton and kinetic energies compared to the intralayer-interlayer exciton splitting $\Delta$ we can readily allow for the exciton dispersion. In that case the $\Delta$ in the denominator in Eq.~\eqref{EXMP} should be replaced by 
\[
\Delta(\bm k, \bm q) = \Delta + \sum_\alpha \left[\frac{\hbar^2(q_\alpha+k_\alpha)^2}{2M^\alpha} - \frac{\hbar^2k_\alpha^2}{2M^\alpha}\right].
\]

The polaron binding energy at $\bm k=0$ reads
\begin{equation}
\label{eq:Exmp-exact}
E_{XMP} = \frac1{\mathcal{N}}\sum_{\boldsymbol q}\mathcal N_{\bm q}\frac{\left|t^e_{\bm q}\right|^2}{4S\Delta(0, \bm q)}.
\end{equation}
Accordingly, the polaron renormalization of the effective mass takes the form
\begin{equation}
    \frac{M^{\alpha}}{M_{XMP}^{\alpha}} \\
    = 1 - \frac{4}{{\mathcal{N}}}\sum_{\boldsymbol q}\mathcal N_{\bm q}\frac{\left|t^e_{\bm q}\right|^2}{4S[\Delta(0, \bm q)]^{{3}}}\frac{\hbar^2q_{\alpha}^2}{2M^{\alpha}},
\end{equation}
where $\alpha\in \{x, y\}$. These expressions are very similar to the expressions for the phonon polaron shift and mass in the weak coupling approach~\cite{doi:10.1080/00018735400101213,Alexandrov2010,PhysRevB.100.041301,2022arXiv220212143I}.

The polaron binding energy and effective mass are shown in Figure~\ref{fig:XMP}. The detailed analytical approximations for exciton-magnon polaron binding energy are given in Appendix~\ref{Appendix:XMP}. The main conclusion here is that these quantities are not large: The exciton-magnon polaron energy does not exceed several~meV for $100$~K and lower for lower temperatures and the mass correction is on the order of fraction of a percent. It means that the magnon-polaron effects can be disregarded in the discussion of the exciton and magnon propagation.
\begin{figure}[ht]
    \centering
    \includegraphics[width=\linewidth]{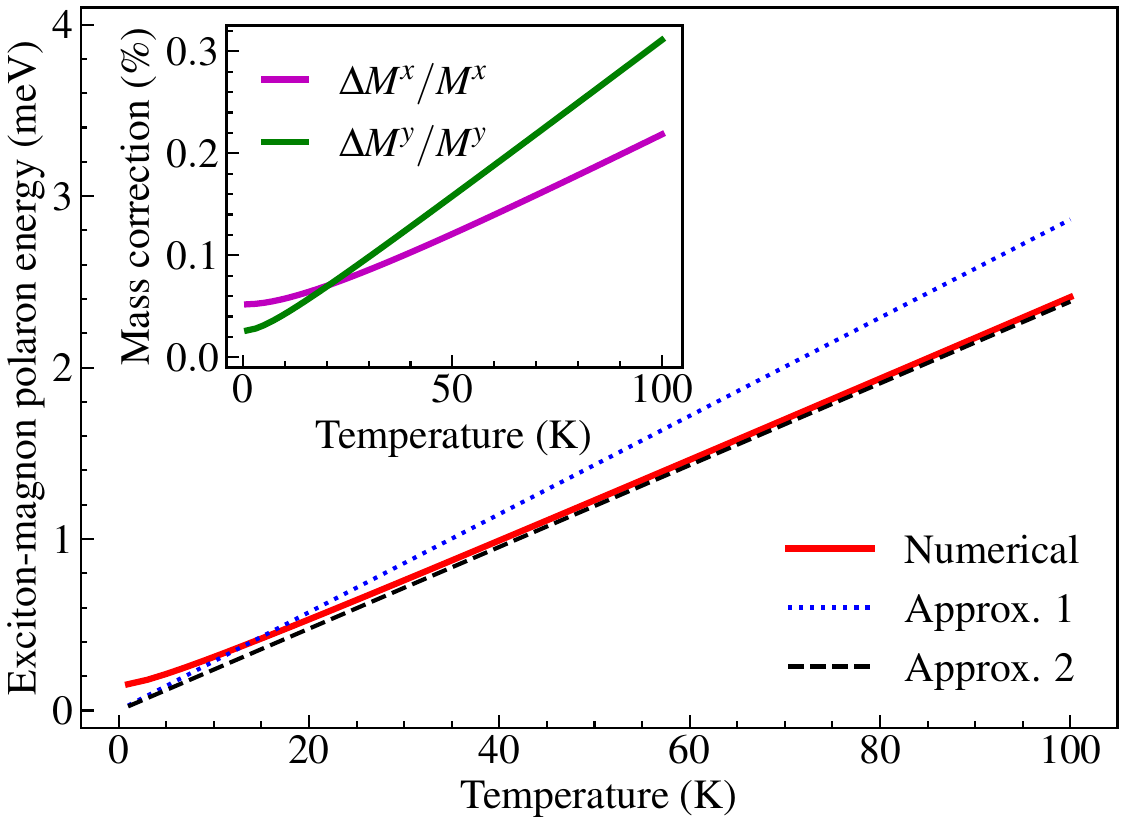}
    \caption{Exciton-magnon polaron binding energy as a function of temperature. Solid red line is numerical calculation, Eq.~\eqref{eq:Exmp-exact}, dotted blue and dashed black lines are analytical approximations (Eq.~\eqref{eq:num1} and Eq.~\eqref{eq:num2}, respectively). Inset shows the relative polaron mass correction ($\Delta M^{\alpha}_{XMP} / M^\alpha = M_{XMP}^\alpha / M^\alpha - 1$) as a function of temperature, $\alpha = x$ (magenta) and $\alpha = y$ (green).}
    \label{fig:XMP}
\end{figure}

\section{Coupled magnon and exciton transport}\label{sec:transport}

\subsection{Kinetic equations and collision integrals}

We now turn to the analysis of propagation of interacting excitons and magnons. We study propagation of the intralayer excitons only. For simplicity, we consider that all kinetic energies are small in comparison to the exciton energy separation~$\Delta$, all momenta are small compared to the exciton inverse Bohr radius. 
For the same reason we consider only electron tunneling between the layers. 
\addZakhar{It allows us to present the relevant part of the Hamiltonian~\eqref{HXM-momentum1} responsible for the exciton transition from the state $\bm k$ to the state $\bm k+ \bm q$ in terms of magnon operators $e_{\bm q}, \ldots, f^\dag_{\bm q}$ as follows:
\begin{multline}
    \label{H:XM:scatt}
    \mathcal{H}_{XM}(\bm k + \bm q, \bm k) = \\ -\frac{1}{\mathcal{S}}\sum_{\bm p}\left[V_{\bm p, \bm p + \bm q}^+\left(2e^\dagger_{\bm p}e_{\bm p + \bm q} + e_{-\bm p}e_{\bm p + \bm q} + e^\dagger_{\bm p}e_{-\bm p - \bm q}\right)\right] \\ -\frac{1}{\mathcal{S}}\sum_{\bm p}\left[V_{\bm p, \bm p + \bm q}^-\left(2f^\dagger_{\bm p}f_{\bm p + \bm q} - f_{-\bm p}f_{\bm p + \bm q} - f^\dagger_{\bm p}f_{-\bm p - \bm q}\right)\right],
\end{multline}
where the matrix elements $V_{\bm p, \bm p + \bm q}^\sigma$ with $\sigma = +$ or $-$ referring to the magnon mode are
\begin{equation}
    V_{\bm p, \bm p + \bm q}^\sigma = \frac{\mathcal{A}_0}{2}\frac{t^{e*}_{\bm p}t^e_{\bm p + \bm q}}{4S\Delta}\frac{X_{\bm p}^{\sigma+} - X_{\bm p}^{\sigma-}}{\sqrt{2}}\frac{X_{\bm p + \bm q}^{\sigma+} - X_{\bm p + \bm q}^{\sigma-}}{\sqrt{2}},
\end{equation}
where the transformation coefficients are discussed in Appendix~\ref{app:Xs}. Note that within our approximations the interactions with magnons in different branches $\sigma =+$ and $-$ are independent.}

\begin{figure}
    \centering
    \includegraphics[width=\linewidth]{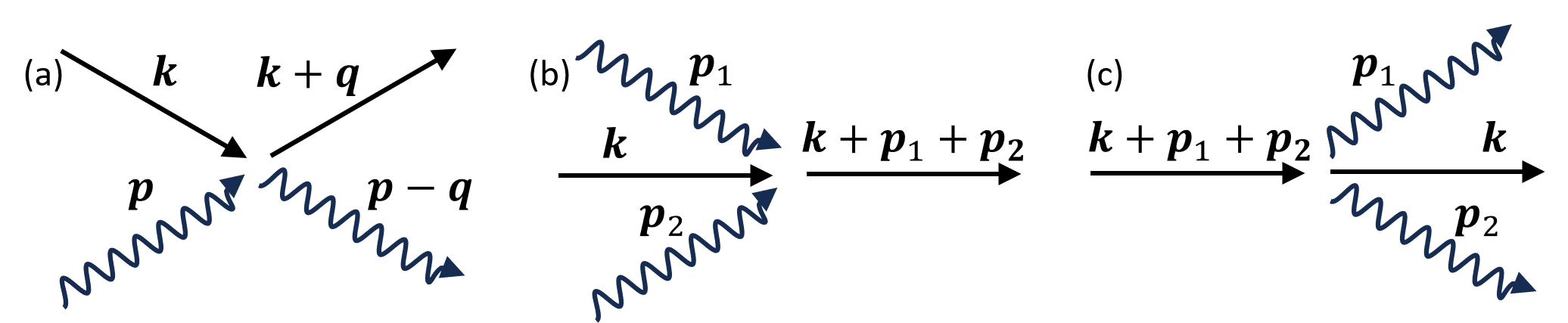}
    \caption{Schematic illustration of the relevant two-magnon processes: (a)~scattering, (b)~two-magnon absorption and (c)~two-magnon emission. \addZakhar{Solid lines represent excitons, while wavy lines represent magnons.}}
    \label{fig:XMint}
\end{figure}

The Hamiltonian is quadratic in the magnon operators, hence, the interaction involves two-magnon processes. It is the specifics of the system in the absence of external magnetic field. Overall, there are three types of two-magnon processes: scattering, two-magnon absorption and two-magnon emission, as depicted in Fig.~\ref{fig:XMint}. \addZakhar{Since the branches are independent, all the processes contain both magnons from the same magnon branch.}

\addZakhar{The characteristic wavevectors for the relevant scattering processes are of the order of thermal wavevectors of excitons and magnons or even smaller for non-equilibrium magnons which are generated optically~\cite{Sun:2024aa}. They are much lower than the inverse Bohr radius and characteristic wavevectors of $X^{\pm\pm}_{\bm k}$ saturation, see Appendix~\ref{app:Xs}. Thus, the matrix elements in this limit are
\begin{equation}
    V_{\bm p, \bm p + \bm q}^\sigma \equiv V^\sigma = \frac{(X^{\sigma+}_0-X^{\sigma-}_0)^2}{2}V = \begin{cases}
        \frac{K_xV}{J_{int} + K_z}, & \sigma = +,\\
        \\
        \frac{K_zV}{J_{int} + K_x}, & \sigma = -,
    \end{cases}
\end{equation}
where we introduced} a constant matrix element of the exciton-magnon interaction
\begin{equation}
    \label{M:x:m}
    V= \frac{\mathcal{A}_0}{2}\frac{|t^e_0|^2}{4S\Delta}.
\end{equation}


The Hamiltonian~\eqref{H:XM:scatt} allows us to derive the kinetic equations for the quasiclassical distribution functions of excitons and magnons: $g^X(\bm k, \bm r, t)$ and $g^\sigma(\bm k, \bm r, t)$. Here $\bm k$ is the wavevector, $\bm r$ is the coordinate, $t$ is time, and the superscript $\sigma=+$ or $-$ denotes corresponding magnon branch. This approach is valid provided that the characteristic length and timescales of the propagation exceed by far the de Broglie wavelengths and frequencies of the involved quasiparticles. 

The kinetic equation for the exciton distribution function can be written as
\begin{equation}
    \frac{\partial g^X}{\partial t} + \bm v_{\bm k}\frac{\partial g^X}{\partial\bm r} = Q^0\left\{g^X\right\} + Q^M\left\{g^X,g^{{+}},g^{{-}}\right\},
    \label{kineq}
\end{equation}
Here, $Q^M$ is the exciton-magnon collision integral describing the processes in Fig.~\ref{fig:XMint} and $Q^0$ is the collision integral responsible for all other scattering processes, e.g., exciton-phonon and exciton-disorder interactions, as well as for the generation and recombination of the excitons. We assume that the exciton population is small, $g^X\ll 1$ \addZakhar{(hence, the exciton-magnon interaction does not affect the magnon distribution function)} and, accordingly, the exciton-magnon collision integral can be written in a linear-in-$g^X$ approximation with two independent magnon modes:
\begin{subequations}
\label{QM}
\begin{equation}
Q^M\left\{g^X,g^{{+}},g^{{-}}\right\} \addZakhar{= Q^+\left\{g^X,g^+\right\} + Q^-\left\{g^X,g^-\right\}},
\end{equation}
\addZakhar{where $Q^\sigma$ refers to the interaction with specific magnon branch}
\begin{multline}
    \addZakhar{Q^\sigma\left\{g^X, g^\sigma\right\} \equiv Q^\sigma_{sc}\left\{g^X,g^{\sigma}\right\}} \\ \addZakhar{+ Q^\sigma_{abs}\left\{g^X,g^{\sigma}\right\} + Q^\sigma_{em}\left\{g^X,g^{\sigma}\right\},}
\end{multline}
\end{subequations}
and the three terms denote the scattering (\emph{sc}), absorption (\emph{abs}), and emission (\emph{em}) processes, respectively:
\begin{subequations}
\label{QM:parts}
  \begin{multline}
  \label{Q:M:sc}
    Q^{\sigma}_{sc}\left\{g^X,g^{\sigma}\right\}  = \frac{2\pi}{\hbar}|2V^{\sigma}|^2 \\\times {\frac{1}{\mathcal{S}^2}}\sum_{\bm p, \bm q}\delta\left(E^X_{\bm k} + E^{\sigma}_{\bm p} - E^X_{\bm k + \bm q} - E^{\sigma}_{\bm p - \bm q}\right) \\\times \left\{g^X(\bm k + \bm q)\left[g^{\sigma}(\bm p) + 1\right]g^{\sigma}(\bm p - \bm q) \right. \\ - \left. g^X(\bm k)g^{\sigma}(\bm p)\left[g^{\sigma}(\bm p - \bm q) + 1\right]\right\},
    \end{multline}
    \begin{multline}
    \label{Q:M:abs}
    Q^{\sigma}_{abs}\left\{g^X,g^{\sigma}\right\}  = \frac{2\pi}{\hbar}|V^{\sigma}|^2
    \\
    \times {\frac{1}{\mathcal{S}^2}}\sum_{\bm p, \bm q}\delta\left(E_{\bm k}^X + E^{\sigma}_{\bm p} + E^{\sigma}_{-\bm p + \bm q} - E^X_{\bm k + \bm q}\right) \\ \times \left\{g^X(\bm k + \bm q)\left[g^{\sigma}(\bm p) + 1\right]\left[g^{\sigma}(-\bm p + \bm q) + 1\right] \right. \\ \left. - g^X(\bm k)g^{\sigma}(\bm p)g^{\sigma}(-\bm p + \bm q)\right\},
    \end{multline}
    \begin{multline}
    \label{Q:M:em}
         Q^{\sigma}_{em}\left\{g^X,g^{\sigma}\right\}  = \frac{2\pi}{\hbar}|V^{\sigma}|^2\\
         \times {\frac{1}{\mathcal{S}^2}}\sum_{\bm p, \bm q}\delta\left(E^X_{\bm k} - E^{\sigma}_{-\bm p} - E^{\sigma}_{\bm p - \bm q} - E^X_{\bm k + \bm q}\right) \\ \times \left\{g^{X}(\bm k + \bm q)g^{\sigma}(-\bm p)g^{\sigma}(\bm p - \bm q) \right. \\ - \left. g^X(\bm k)\left[g^{\sigma}(-\bm p) + 1\right]\left[g^{\sigma}(\bm p - \bm q) + 1\right]\right\}.
\end{multline}
\end{subequations}
Note that in scattering collision integral~\eqref{Q:M:sc} the matrix element $2V$ is doubled \addZakhar{due to the factor of 2} in Hamiltonian~\eqref{H:XM:scatt}. 
Similar kinetic equation and collision integrals can be written for magnons.

\subsection{Distribution functions}

We assume for simplicity that the lifetimes of the quasiparticles, excitons and magnons, are sufficiently long and neglect the generation and recombination processes. Let us consider the model situation where the magnon distribution function is determined by the excitation conditions. To illustrate the approach we take it in a quasi-equilibrium form that corresponds to the propagation of the magnon cloud with the velocity $\bm u$
\begin{equation}
\label{magnon:distrib}
    g^{\sigma}(\bm p) = \frac{1}{\exp\left(\frac{E^{\sigma}_{\bm p} - \hbar\bm p\bm u}{k_BT}\right) - 1} \approx g^{\sigma}_0(\bm p) + \delta g^\sigma(\bm p),
\end{equation}
with 
\begin{equation}
\label{delta:g:m}
    \delta g^\sigma(\bm p) = -\frac{\partial g_0^\sigma(\bm p)}{\partial E^\sigma_{\bm p}}\hbar\bm p\bm u = g^{\sigma}_0(\bm p)\left[g_0^{\sigma}(\bm p) + 1\right]\frac{\hbar\bm p\bm u}{k_BT}.
\end{equation}
Here the drift velocity $\bm u$ is related to the magnon flow, the function $g^{\sigma}_0(\bm p)$ is $g^\sigma(\bm p)$ at $\bm u=0$ \addZakhar{(Planck's distribution)}. In this part, we consider only the case where the thermal magnon flow is induced by a weak perturbation, e.g., by temperature gradient. The effects of nonequilibrium magnon distribution and excitation are discussed below in Sec.~\ref{subsec:drag}. We assume that $|\delta g^\sigma(\bm p)|\ll g_0^\sigma(\bm p)$ and linearize collision integral with respect to $\delta g^\sigma(\bm p)$.

Similarly, the exciton distribution is also taken in the quasi-equilibrium form
\begin{equation}
    \label{x:distrib}
g^X(\bm k) = g^X_0(\bm k)\left(1+ \frac{\hbar \bm k \bm v}{k_B T}\right),
\end{equation}
where $g^X_0(\bm k) = \exp{[(\mu^X-E_{\bm k}^X)/k_B T]}$ is the Boltzmann distribution function with the same temperature as that of magnons. The chemical potential $\mu^X$ is determined by the exciton density and the exciton drift velocity $\bm v$ is to be found from the kinetic equation. As already mentioned, since exciton density is assumed to be small, the effect of excitons on magnon propagation and magnon distribution can be neglected\addZakhar{; the analysis of this approximation is presented in the end of Sec.~\ref{subsec:drag}}.

\subsection{Collision rates}

The collision integral $Q^{\sigma}$, Eq.~\eqref{QM}, vanishes if both exciton and magnon distribution functions are taken in the zero-order approximation:
\[
Q^{\sigma}\{g^X_0,g^{\sigma}_0\}=0.
\]
The linear in $\delta g^{\sigma}$ and $\delta g^X$ contributions, $Q^{\sigma}\{g^X_0,g_0^{\sigma}+\delta g^{\sigma}\}$ and $Q^{\sigma}\{\delta g^X,g^{\sigma}_0\}$, describe the magnon-induced generation of exciton flux and its relaxation, respectively. \addZakhar{The detailed derivation of collision integrals for all types of the exciton-magnon interaction processes is provided in Appendix~\ref{app:collision}. Here we summarize the key results.}

\emph{Scattering processes.} \addZakhar{Derivation presented in Appendix~\ref{app:collision:sc} shows} the scattering part of the collision integral takes a very simple form: 
\begin{equation}
    Q^{\sigma}_{sc}\left\{g_0^X,g^{\sigma}\right\} = 2\pi|2V^{\sigma}|^2\mathcal D^M\mathcal D^X \frac{k_BT}{E_0^{\sigma}}g^X_{0}(\bm k)\addZakhar{\nu_{\alpha\beta}(\bm k)k_{\alpha}u_{\beta}},
\end{equation}
where magnon and exciton densities of states are
\[
\mathcal{D}^M = \frac{\sqrt{m_m^xm_m^y}}{2\pi\hbar^2}, \qquad \mathcal{D}^X = \frac{\sqrt{M^XM^Y}}{2\pi\hbar^2},
\]
and $E_0^{\sigma} \equiv E^\pm_{\bm k=0} \sim  0.1~\text{meV} \ll k_BT$ is magnon cut-off energy \addZakhar{different but close for two branches. The cut-off energy appears in Eq.~\eqref{eq:Qsc} due to the constant density of states in the two-dimensional system. We also recall that we consider equilibrium magnons with zero chemical potential, see Eq.~\eqref{magnon:distrib}. For nonthermal magnons the cut-off energy could be related to the absolute value of the chemical potential. Here $\nu_{\alpha\beta}(\bm k)$ is almost isotropic diagonal tensor close to unity $2\times 2$ matrix, see Appendix~\ref{app:collision:sc} and Fig.~\ref{fig:nu} in particular. At small exciton energies
\begin{equation}
\label{condition:iso}
E^X_{\bm k} \sim k_BT \lesssim E_0^{\sigma}\sqrt{\frac{m_m^x m_m^y}{M^xM^y}},
\end{equation}
where $m_m^x,m_m^y$ are the effective magnon masses, the scattering becomes isotropic
\begin{equation}
    \label{eq:Qsc}
    Q^{\sigma}_{sc}\left\{g_0^X,g^{\sigma}\right\} = \addZakhar{2}\pi|2V^{\sigma}|^2\mathcal D^M\mathcal D^X \frac{k_BT}{E_0^{\sigma}}g^X_{0}(\bm k)\bm k\bm u.
\end{equation}
} 

Performing similar derivations to obtain $Q^{\sigma}_{sc}\{\delta g^X,g_0^{\sigma}\}$ we get
\begin{equation}
\label{Q:scatt:2}
    Q^{\sigma}_{sc}\left\{g^X,g^{\sigma}\right\} = \addZakhar{2}\pi|2V^{\sigma}|^2\mathcal D^M\mathcal D^X
    \frac{k_BT}{E_0^{\sigma}}g^X_0(\bm k)\bm k(\bm u-\bm v).
\end{equation}

\emph{Two-magnon absorption \addZakhar{and emission}.} \addZakhar{Detailed derivation of the contributions of these processes to the collision integral is presented in Appendixes~\ref{app:abs} and \ref{app:em}. It turns out that
\begin{equation}
    \label{Q:em:2}
    Q^\sigma_{em}\left\{g^X, g^\sigma\right\} = Q^\sigma_{abs}\left\{g^X, g^\sigma\right\},
\end{equation}
and}
\begin{equation}
\label{Q:abs:2}
    Q^{\sigma}_{abs}\left\{g^X,g^{\sigma}\right\} = 2\pi|V^{\sigma}|^2\mathcal D^M\mathcal D^X\frac{k_BT}{E_0^{\sigma}}g^X_0(\bm k)\bm k(\bm u-\bm v),
\end{equation}
\addZakhar{where for compactness we present the small-energy expression (full expression is given in Eqs.~\eqref{Q:abs:1}).} Note that the emission and absorption processes provide the same contributions to the collision integral that are 4 times smaller, than the scattering processes.

Equations~\eqref{Q:scatt:2}, \eqref{Q:abs:2}, and \eqref{Q:em:2} make it possible to recast the exciton-magnon collision integral (with allowance for all processes depicted in Fig.~\ref{fig:XMint}) in the relaxation time approximation
\begin{equation}
\label{collision:relax:time}
    Q^{\sigma}\{g^X,g^{\sigma}\} = \frac{1}{\tau^{\sigma}_{XM,\alpha\beta}} g_0^X(\bm k) \frac{\hbar \addZakhar{k_{\alpha} (u_\beta - v_\beta)}}{k_B T},
\end{equation}
where the exciton-magnon scattering time is given by
\begin{subequations}
    \label{relaxation:time}
    \begin{equation}
        \label{eq:rate:num}
        \frac{1}{\tau^{\sigma}_{XM\addZakhar{,\alpha\beta}}} = 12\pi|V^{\sigma}|^2\mathcal D^M\mathcal D^X\frac{k_B T}{\hbar}\addZakhar{\frac{k_BT}{E^\sigma_0}\nu_{\alpha\beta}(\bm k)},
    \end{equation}
\addZakhar{Under condition~\eqref{condition:iso} we have}
    \begin{equation}
        \frac{1}{\tau^{\sigma}_{XM,\alpha\beta}} \equiv \frac{1}{\tau^\sigma_{XM}} = 12\pi|V^\sigma|^2\mathcal D^M\mathcal D^X\frac{k_B T}{\hbar}\frac{k_BT}{E^\sigma_0}.
    \end{equation}
\end{subequations}
The collision integral~\eqref{collision:relax:time} describes the relaxation of exciton and magnon distribution functions towards the distributions that correspond to equal propagation velocities. It describes the drag of excitons by magnons.

\addZakhar{The total exciton-magnon scattering rate is the sum of scattering rates over the magnon branches, that is for both types of magnons moving with the same velocity is
\begin{multline}
    \label{TauXM:exact}
    \frac{1}{\tau_{XM,\alpha\beta}} = \frac{1}{\tau^+_{XM,\alpha\beta}} + \frac{1}{\tau^-_{XM,\alpha\beta}} \\ = 12\pi\mathcal{D}^M\mathcal{D}^X\frac{(k_BT)^2}{\hbar}\left(\frac{|V^+|^2}{E^+_0} + \frac{|V^-|^2}{E^-_0}\right)\nu_{\alpha\beta}(\bm k).
\end{multline}
}

\begin{figure}[ht]
    \centering
    \includegraphics[width=\linewidth]{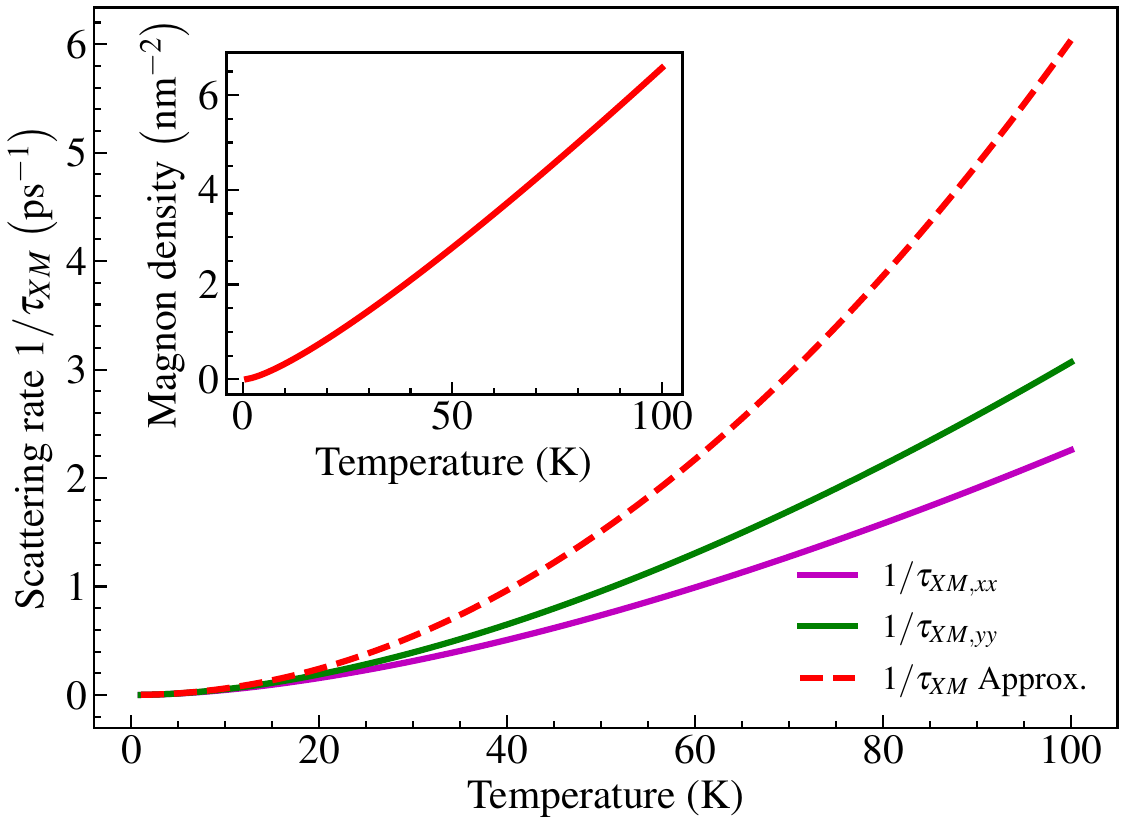}
    \caption{Exciton-magnon scattering rate as a function of temperature. \addZakhar{Solid magenta and green lines are exact exciton-magnon scattering rates for excitons moving along $x$- and $y$- directions respectively, Eq.~\eqref{TauXM:exact}, dashed red line is an approximate expression~\eqref{eq:rate:approx} valid for small temperatures.} 
    Inset shows the magnon density as a function of temperature~\eqref{eq:magnon_density} for $E_0^M = 0.1$~meV.}
    \label{fig:tau}
\end{figure}

Exciton-magnon scattering rate averaged over Boltzmann distribution of excitons as a function of temperature together with its asymptotic isotropic approximation
\begin{multline}
    \label{eq:rate:approx}
    \frac{1}{\tau_{XM}} = 12\pi|V|^2\mathcal D^M\mathcal D^X\frac{(k_B T)^2}{\hbar}\\ \addZakhar{\times \left[\frac{\sqrt{K_x}}{2S(J_{int} + K_z)^{3/2}} + \frac{\sqrt{K_z}}{2S(J_{int} + K_x)^{3/2}}\right]},
\end{multline}
is plotted in Fig.~\ref{fig:tau}\addZakhar{, where the expression in square brackets is from the terms $|V^\sigma|^2/E^\sigma_0$. Note that $1/\tau^-_{XM}$ larger than $1/\tau^+_{XM}$ in a factor of
\begin{equation}
    \label{eq:tau:ratio}
    \frac{\tau^+_{XM}}{\tau^-_{XM}} = \sqrt{\frac{K_z}{K_x}}\left(\frac{J_{int} + K_z}{J_{int} + K_x}\right)^{3/2} \sim 12
\end{equation}
for bilayers. Note that the results are sensitive to the parameters, presented in Table~\ref{tab:param}.} Scattering rate increases quadratically with temperature as it is induced by two-magnon processes. The inset to the figure shows the thermal magnon density as a function of temperature:
\begin{multline}
    \label{eq:magnon_density}
    N_{M} = \frac{2}{\mathcal S}\sum_p g^M_{0}(\bm p) = -2\mathcal{D}^Mk_BT\ln\left(1 - e^{-\frac{E_0^M}{k_BT}}\right)\\
    \approx 2\mathcal{D}^Mk_BT\ln\frac{k_BT}{E_0^M},
\end{multline}
\addZakhar{where for simplicity we disregarded differences in $E^+_{\bm p}$ and $E^-_{\bm p}$.}
It is important to note that the exciton-magnon scattering time falls into subpicosecond range for the moderate temperatures of few tens of Kelvin which are significantly smaller than the N\'eel temperature in CrSBr. It means that exciton interaction even with equilibrium magnons can be of paramount importance for exciton momentum relaxation and propagation as we discuss below.

\addZakhar{For completeness, we present an estimate of the exciton-phonon momentum relaxation rate extending the well-known approaches~\cite{kaasbjergPhononlimitedMobilityNtype2012,shreeObservationExcitonphononCoupling2018,10.1063/5.0122633} to allow for the anisotropy of the exciton dispersion
\begin{equation}
    \frac{1}{\tau_{ph}} = \frac{k_BT \sqrt{M_xM_y}\Xi^2}{\hbar^3\rho s_{LA}^2},
\end{equation}
where $\Xi \approx 3$~eV is the averaged deformation potential~\cite{D3NR02518G}, $\rho = 3.2\cdot 10^{-7}~\text{g/cm}^2$ is two-dimensional CrSBr mass density, and $s_{LA} = 5.2$~km/s is the averaged longitudinal phonon speed~\cite{10.1063/4.0000266}. We account for the longitudinal acoustic phonons only because this mechanism is known to be the most efficient one at not too high temperatures~\cite{Chernikov:2023ab}. At $T = 100$~K the phonon-induced momentum relaxation rate is $\tau_{ph}^{-1} = 7$~ps$^{-1}$, i.e., it is comparable and even somewhat larger than the exciton-thermal magnon scattering rate. Interestingly, that the two mechanisms provide different temperature dependence of the scattering rate: $\propto T$ for acoustic phonons and $\propto T^2$ for thermal magnons. In experiment~\cite{crsbr:exp}, as we discuss below, the exciton scattering and drag is provided by non-equilibrium magnons. As for the static disorder scattering, it is expected to be weakly temperature dependent, its rate can be roughly estimated from the linewidth broadening at $T=0$ yielding $1/\tau_{dis} \lesssim 1$~ps$^{-1}$~\cite{crsbr:exp}.}

\section{Magnon-exciton drag}\label{sec:drag}

\subsection{Exciton propagation}\label{subsec:propagation}

The kinetic equation and scattering rates derived above provide a general framework for the description of exciton and magnon propagation. Let us now establish a relation between the developed model and the results of Ref.~\cite{crsbr:exp} where a number of unusual effects in exciton transport in CrSBr had been observed. Making use of the general kinetic Eq.~\eqref{kineq} and the collision integral~\eqref{collision:relax:time} we obtain the following equation for the exciton ensemble velocity $\bm v$
\begin{equation}
    \label{equation:eq:v}
    \frac{d\bm v}{dt} + \hat\Gamma \bm v + \frac{\bm v - \bm u}{\bar\tau_{XM}} = \bm f,
\end{equation}
where
\[
\frac{1}{\bar\tau_{XM}} = \frac{\sum_{\bm k} \tau_{XM}^{-1} E_{\bm k}^X g_0^X(\bm k)}{\sum_{\bm k}  E_{\bm k}^X g_0^X(\bm k)}
\]
is the averaged exciton-magnon scattering time\footnote{Here the averaging assumes ``hydrodynamic'' form of the transport~\cite{gantmakher87,mantsevich2024viscoushydrodynamicsexcitonsvan}. Note that $\tau_{XM}$ in Eq.~\eqref{relaxation:time} just weakly depends on the energy and the way of averaging is, in fact, unimportant.} and $\hat\Gamma$ is the rank-2 tensor of exciton relaxation rates unrelated to the scattering by magnons. \addZakhar{Hereafter we neglect the small anisotropy in $\tau_{XM,\alpha\beta}$.} The right-hand side in Eq.~\eqref{equation:eq:v} contains the effective force $\bm f$ (divided by corresponding effective masses along the principal axes) related to the perturbations acting on excitons. In derivation of Eq.~\eqref{equation:eq:v} we assumed that variations of exciton density
\begin{equation}
    N_X=\mathcal S^{-1}\sum_{\bm k} g_0^X(\bm k),
\end{equation}
occur on a time scale being much longer than $\bar\tau_{XM}$ and $\hat{\Gamma}^{-1}$ and neglected the terms responsible for the exciton viscosity, the latter is possible if exciton density is not too high~\cite{mantsevich2024viscoushydrodynamicsexcitonsvan}.

The exciton density as a function of coordinate $\bm r$ and time $t$ obeys the continuity equation in the form
\begin{equation}
    \label{equation:continuity:x}
    \frac{\partial N_X}{\partial t} + \frac{\partial}{\partial \bm r}(\bm v N_X) + \frac{N_X}{\tau_r}=0, 
\end{equation}
where $\tau_r$ is the exciton recombination time and we disregard for simplicity nonlinear recombination processes. The set of Eqs.~\eqref{equation:eq:v} and \eqref{equation:continuity:x} together with expressions for $\bm u$ and $\bm f$ (see below) describes the exciton propagation on the time scales which exceed by far the microscopic scales related to the $\hbar/E_{\bm k}^X$ and $\bar \tau_{XM}$, see Ref.~\cite{Chernikov:2023ab} for details on the applicability of this approach.

\subsection{Exciton diffusion and drag by magnons}\label{subsec:drag}

Let us now discuss two key regimes of exciton propagation which can be expected in the studied system. First, \emph{diffusive} propagation regime naturally arises where magnon propagation can be neglected, i.e., at $\bm u=0$. Such a regime may be realized at weak quasi-equilibrium optical excitation which generates a small density of excitons without affecting the magnetic subsystem. The steady-state exciton velocity (at $t \gg \hat\Gamma^{-1}, \bar\tau_{XM}$) reads
\begin{subequations}
    \label{diffusion}
\begin{equation}
    \bm v = \left(\hat \Gamma + \frac{1}{\bar\tau_{XM}}\right)^{-1}\bm f,
\end{equation}
and ($\alpha=x$ or $y$)
\begin{equation}
    f_\alpha = - \frac{1}{M^\alpha} \frac{\partial \mu^X}{\partial x_\alpha},
\end{equation}
with $\mu^X$ being exciton chemical potential. Equation~\eqref{equation:continuity:x} then reduces to the classical diffusion equation 
\begin{equation}
    \label{equation:diffusion:x}
    \frac{\partial N_X}{\partial t} + D_\alpha \frac{\partial^2}{\partial x_\alpha^2} N_X + \frac{N_X}{\tau_r}=0, 
\end{equation}
with the diffusion coefficients along the principal axes of the structure in the form
\begin{equation}
\label{D:alpha}
D_\alpha = \frac{k_B T}{M^\alpha}  \left(\Gamma_{\alpha\alpha} + \frac{1}{\bar\tau_{XM}}\right)^{-1}.
\end{equation}
\end{subequations}
In this regime exciton diffusion is expected to be strongly anisotropic because of significant difference of the exciton effective masses $M^x$ and $M^y$. \addZakhar{Estimated diffusion coefficients are $D_y \sim 1$~cm$^2/$s and $D_x \sim 0.05$~cm$^2$/s at low temperatures based on the values of $\Gamma$ extracted from the experimental linewidths~\cite{crsbr:exp}, see also Ref.~\cite{wagnerNonclassicalExcitonDiffusion2021}.} Moreover, one can expect that $D_\alpha$ drops as $1/T$ with increase in the temperature: Indeed, the exciton-magnon scattering time $\tau_{XM}$ decreases as $1/T^2$, see Eq.~\eqref{eq:rate:approx} and Fig.~\ref{fig:tau} while the prefactor in the diffusion coefficient $D_\alpha$ increases only linearly with the temperature reflecting the rise of excitons kinetic energy. It is natural for exciton scattering by magnetic fluctuations: the higher the temperature is, the stronger are the fluctuations. Similar effects were studied in the context of electron conductivity in magnetic systems with emphasis on the effects in the vicinity of the Curie or N\'eel temperature where the fluctuations are particularly strong and critical phenomena are of importance~\cite{De-Gennes:1958aa,PhysRevLett.20.665,Suezaki:1969aa,KASUYA1974253,Ausloos:1977aa}.

However, such diffusive behavior is not generally realized in the state-of-the-art experiments~\cite{crsbr:exp}. Thus we turn to the second regime of \emph{exciton drag by magnons} which is expected to occur if significant non-equilibrium magnon populations are optically injected. The magnons are known to propagate for significant distances in CrSBr~\cite{Sun:2024aa} with their fluxes being induced by optical excitation, temperature and density gradients, cf. Refs.~\cite{Adachi_2013,Cui:2024aa}. The non-equilibrium flux of magnons gives rise, as a result of exciton-magnon interactions, to flux of excitons, see Eq.~\eqref{equation:eq:v}. As a result, excitons are dragged by magnons similarly to the electron-magnon drag~\cite{Flebus_2016,PhysRevB.99.094425,PhysRevB.105.045101}. These effects are similar to the phonon drag and wind effect in metals and semiconductors~\cite{legurevich:drag,GUREVICH1989327,keldysh_wind,PhysRevB.100.045426}.

It follows from Eq.~\eqref{equation:eq:v} that the efficiency of the magnon drag effect is controlled by the dimensionless parameters $\bar\tau_{XM}\Gamma_{\alpha\alpha} $ which describe (at $\bm f=0$) the ratio of the exciton momentum loss rate by magnon-unrelated mechanisms (principal components $\Gamma_{\alpha\alpha}$ of the tensor $\hat \Gamma$) and exciton-magnon momentum exchange rate $1/\bar\tau_{XM}$. The exciton drift velocity is given by 
\begin{equation}
    \label{exc:drift}
    \bm v = \left(1 + \bar \tau_{XM}\hat \Gamma\right)^{-1} \bm u.
\end{equation}
As mentioned above and depicted in Fig.~\ref{fig:tau} even for equilibrium or quasi-equilibrium magnons the scattering rates exceed units of ps$^{-1}$ for moderate temperatures making it plausible that exciton-magnon scattering dominates over other relaxation processes. Under non-equilibrium conditions the scattering rate roughly scales as $N_M^2$ since two-magnon processes are important. For instance, in the case of non-resonant optical excitation we can envisage the following scenario:

(i) optically excited non-equilibrium excitons and electron-hole pairs lose their energy and generate magnons within the excitation spot, additionally magnons can be generated directly by the optical pulse~\cite{Kimel:2020aa,KalashnikovaAM2022,2025arXiv250702793V};

(ii) the magnons relax in energy and form a quasiequilibrium distribution similar to Eq.~\eqref{magnon:distrib} but, generally, with non-zero chemical potentials $\mu^\sigma$ and effective temperature $T^M$:
\begin{equation}
\label{magnon:distrib:non:eq}
    g^{\sigma}(\bm p) = \frac{1}{\exp\left(\frac{E^{\sigma}_{\bm p} - \hbar\bm p\bm u - \mu^\sigma}{k_BT^M}\right) - 1};
\end{equation}

(iii) a non-equilibrium flux of magnons with (generally, coordinate and time dependent velocity) $\bm u$ is formed.

These non-equilibrium magnons drag excitons. Note that for the magnon distribution in the form of Eq.~\eqref{magnon:distrib:non:eq} the exciton-magnon scattering rates can be expressed in a form similar to Eqs.~\eqref{collision:relax:time} and \eqref{eq:rate:approx} with different cut-off energy and temperature. For instance, neglecting the difference between the magnon branches we obtain, instead of Eq.~\eqref{eq:rate:approx}, an estimate
\begin{equation}
    \label{eq:rate:approx:non:eq}
    \frac{1}{\tau_{XM}^{\rm neq}} = 48\pi|V|^2\mathcal D^M\mathcal D^X\frac{k_B T}{\hbar}\frac{k_BT^M}{E_0^M-\mu^M}.
\end{equation}
\addZakhar{Note that if the density of non-equilibrium magnons is high enough, the scattering by non-equilibrium magnons dominates over the scattering by thermal magnons, acoustic phonons, and static disorder:  $\tau_{XM}^{\rm neq} \ll \tau_{XM}, \tau_{ph}$ and in Eq.~\eqref{exc:drift} $\bar \tau_{XM} \equiv \tau_{XM}^{\rm neq}$.}


Hence, at $\bar\tau_{XM}\Gamma_{\alpha\alpha} \ll 1$ the excitons and magnons propagate with the same velocity $\bm u$ determined by the magnon excitation conditions because, at least for small exciton densities, the feedback of excitons on magnon propagation is negligible. Such regime offers unique propagation features of excitons such as: (i) reduced anisotropy of exciton propagation compared to the classical diffusion regime because of less pronounced anisotropy of magnon dispersion [cf. Fig.~\ref{fig:disp} \addZakhar{and Fig.~\ref{fig:tau}}] compared to bare exciton dispersion~\cite{Klein:2023,Wilson:2021aa}; and (ii) enhanced exciton propagation velocities, particularly, in the vicinity of the N\'eel temperature where the magnonic spin excitations are strongly populated and exciton-magnon scattering is particularly significant. These features are in line with recent experimental observations in Ref.~\cite{crsbr:exp} supporting the exciton drag by magnons scenario realized in few layer CrSBr van der Waals crystals. We refrain from detailed quantitative comparison with experiments which requires microscopic calculation of the magnon distribution function and velocity $\bm u$ under experimental conditions, which are beyond the scope of present work.

\addZakhar{The scenario outlined above is valid at sufficiently low exciton densities. First, the condition $N_X \ll N_M$ should hold. This condition allows us to disregard the exciton back-action on magnons and use the set of two Eqs. \eqref{equation:eq:v}, \eqref{equation:continuity:x} to describe exciton dynamics at a given magnon propagation velocity $\bm u$. Otherwise, additional equations for the magnon velocity $\bm u$ and magnon population $N_M$ in the form similar to Eqs. \eqref{equation:eq:v}, \eqref{equation:continuity:x}  should be included. Based on our estimates of thermal magnon density on the order of $1$~nm$^{-2}$  (see inset in Fig.~\ref{fig:tau}) the effects of excitons on magnon dynamics can be neglected for typical exciton densities below $10^{14}$ cm$^{-2}$. According to Ref.~\cite{crsbr:exp}, it is indeed the case in experiments even at the largest fluences of $10^4$~$\mu$J/cm$^2$.}

\addZakhar{Note that at large fluences the elevated exciton densities result in  exciton-exciton interactions that produce two additional density-dependent effects: non-linear Auger-like recombination and exciton-exciton repulsion. Both effects are known to increase effective diffusion coefficient of excitons with increase in the exciton density, see Refs.~\cite{Chernikov:2023ab,wietekNonlinearNegativeEffective2024}. The values of the exciton-exciton interaction parameters for CrSBr are not precisely established to the best of our knowledge, but based on the estimates of the Auger recombination rate in Ref.~\cite{crsbr:exp} and crude estimates of the exciton-exciton repulsion we provide an upper limit for exciton density $N_X < 10^{13}$ cm$^{-2}$ (at temperatures below the N\'eel temperature) where the non-linear effects can be neglected. Such a high exciton density limit is related to small Bohr radii of excitons in CrSBr owing to relatively large effective masses of the charge carriers.}

\section{Discussion and outlook}\label{sec:discussion}
We now discuss some limitations of our theoretical method and future directions to address them. First, we have assumed quasiequilibrium distributions for the excitations, which is valid for only a fraction of the experimentally realized situations and phenomena. The issue of determining the true nonequilibrium distributions of quasiparticles on optical excitation remains a hard problem to address. To obtain such a distribution, one needs to know all the excitations present in the solid together with their mutual interactions. Assuming this complete information, one needs to be able to simulate the entire energy transfer process from the optical pump to the different excitations. Alternately, one can assume, phenomenologically and based on some general arguments, excitation distributions parametrized by a few variables that are far-from-equilibrium but are not evaluated rigorously. The next step is to evaluate physical observables and compare them to the experimental data thereby ascertaining the variables parameterizing the assumed distributions. This approach can capture phenomena forbidden in the quasiequilibrium theory and offer an experimental-data-informed phenomenological alternative to far-from-equilibrium phenomena.

Along the same lines, our quasiequilibrium analysis does not capture \addZakhar{two interesting effects observed in experiments~\cite{crsbr:exp} at liquid helium temperature: (i) ultrafast superdiffusive exciton expansion in bilayer and (ii) the negative magnon-exciton drag in a 10-layer sample. Qualitatively, physics of both effects can be explained in terms of magnon-exciton drag by highly non-equilibrium long-wavelength magnons. In bilayer crystals the interaction with the top magnon branch with positive group velocity is dominant, Eq.~\eqref{eq:tau:ratio}, and the drag by expanding top branch is dominant. In contrast, in multilayer crystals the effective $J_{int}$ is enhanced due to the presence of two neighboring layers instead of one and, more importantly, the effects of dipole-dipole interaction and negative group velocity are scaled linearly with the number of layers, see Appendix~\ref{app:FiniteThickness}}. Negatively dispersing, due to dipolar interactions, magnons have group velocity opposite to their wavevector or momentum. While the excitation propagation is governed by the group velocity, the drag effect is governed by a transfer of momentum. Thus, under a highly nonequilibrium distribution function for magnons such that the negatively dispersing low-wavevector modes are predominantly occupied, the magnons would appear to move in a direction opposite to the one along which they drag the excitons resulting in a negative drag effect. Such an effect provides a negative contribution to the exciton diffusivity and is consistent with the exciton cloud contraction observed in recent experiments~\cite{crsbr:exp}. 

Going beyond the distribution functions, let us examine the limitations of our Boltzmann transport theory which treats the excitations as particle-like wave packets while evaluating their collision integrals assuming an infinite-extent plane wave character. As a consequence, in this framework, the exciton drag velocity can only be as large as the magnon drift at maximum [Eq.~\eqref{exc:drift}]. This can be understood within a classical particle-like picture --- the particle cloud being dragged cannot go faster than the particle cloud dragging it along. On the other hand, in a quantum plane wave-like picture which considers the drag to be a result of momentum transfer from one infinite-extent wave-like excitation cloud to the other, the amount of momentum transfer and thus imparted drag velocity is only limited by the density of the excitation cloud. This is the essence of theoretically divergent drag velocities evaluated employing the Green's function framework in the context of magnon-electron drag in antiferromagnets close to their ordering temperature~\cite{Sugihara:1973}. Thus, a future study of transport using Green's function could account for more wave-like properties and predict phase-coherent effects that are yet to be observed experimentally.

\section{Conclusion}\label{sec:conclusion}


We have developed a theory of intertwined exciton-magnon transport in bilayers of van der Waals semiconductor CrSBr. We have put forward the microscopic model of the magnon dispersion in this system that takes into account short-range ferro- and antiferromagnetic exchange interactions between Cr spins, as well as the long-range dipole-dipole interaction. The latter gives rise to the anomalous magnon dispersion with negative group velocity near the Brillouin zone center.
%
We have developed a model of the orbital mechanism 
of exciton-magnon interaction specific to layered antiferromagnets. In our model, magnons tilt 
the  layer magnetizations and, correspondingly, enable tunneling of the charge carriers and excitons between the layers which is spin-forbidden in the perfectly antiferromagnetic configuration. Consequently, magnons mix intra- and interlayer excitons, and modulate the exciton energy. 
We have shown that the interaction arising from the tunneling of charge carriers between neighboring monolayers appears in the second order of perturbation and has a two-magnon nature in accordance with the symmetry analysis.

We have studied two main consequences of the exciton magnon interaction. First, we have analyzed the effect of the interactions on exciton energy and dispersion, i.e., the exciton-magnon polaron effect. The estimations show that the polaron binding energy is $\lesssim 1$~meV and the enhancement of the effective mass is in $\sim 1\%$ range for realistic parameters. Second, we have developed microscopic theory of the exciton-magnon relaxation processes stemming from the two-magnon absorption, two-magnon emission, and scattering. Overall exciton magnon scattering rate calculated within our model grows quadratically with the increase in the temperature and reaches sub-picosecond timescales at several tens to hundred K. Consequently, these efficient relaxation processes can dominate other relaxation mechanisms and are of particular importance for the description of exciton propagation and transport. We have demonstrated that under realistic conditions excitons and magnons can co-propagate with similar velocities, because of the effective momentum exchange, i.e., the magnon-exciton drag can be realized. As a result, the exciton propagation can be strongly enhanced and, depending on the magnon excitation and propagation conditions result
both in negative diffusion or super-diffusion, depending on the excitation conditions. 

Our results provide a theoretical basis for recent experimental observations of anomalous exciton transport in CrSBr and highlight the potential of magnon-driven exciton propagation in layered magnetic semiconductors. Future work can be focused on microscopic modeling of non-equilibrium magnon distributions and exciton-magnon collective behavior both in absence and presence of the external magnetic field providing background for studies of further effects including non-classical and potentially nonreciprocal transport as well as the effects of magnon condensation on exciton propagation.


\section*{Acknowledgment}
We thank F.~Dirnberger, A.~Chernikov, D.~Erkensten and E.~Malic for helpful discussions. Z.A.I. and M.M.G are grateful to RSF Project 23-12-00142 for financial support. Z.A.I gratefully acknowledges the BASIS foundation. 

\bibliography{mag_exc_drag}

\newpage

\appendix

\section{Magnon dispersion in semiclassical approach}
\label{app:semiclass}

The magnetic energy density of bilayer can be expressed in terms of quasicontinuous spin density~$\bm S(\bm r)$ (analogue of Eq.~\eqref{H:magnetic:tot}, $\mathcal{H}_M = \int w(\bm r)d \bm r$) in the most general form as
\begin{equation}
    \label{eq:w(r)}
    w_{M}(\bm r) = w^{FM}(\bm r) +  w^{AFM}(\bm r) + w^a(\bm r) + w^h(\bm r),
\end{equation}
where ferromagnetic interlayer term is
\begin{multline}
    w^{FM}(\bm r) = - \frac{\A}{2}\left[\bm S^{(1)}(\bm r)\cdot \sum_{n'}J_{FM}(n')\bm S^{(1)}(n') \right. \\ + \left. \bm S^{(2)}(\bm r)\cdot \sum_{n'}J_{FM}(n')\bm S^{(2)}(n')\right],
\end{multline}
where $\A = \mathcal{A}_0 / 2$ is the magnetic unit cell (area per one spin), summation over $n'$ is summation over all other magnetic unit cells and $1/2$ for considering each pair of spins once.

Antiferromagnetic interlayer energy density is
\begin{equation}
    w^{AFM}(\bm r) = J_{int}\A \bm S^{(1)}(\bm r)\cdot\bm S^{(2)}(\bm r),
\end{equation}
and the anisotropic component is
\begin{multline}
    w^a(\bm r) = K_x\A\left[S^{(1)}_x(\bm r)^2 + S^{(2)}_x(\bm r)^2\right] \\ + K_z\A\left[S^{(1)}_z(\bm r)^2 + S^{(2)}_z(\bm r)^2\right].
\end{multline}
The interaction energy with magnetic field $\bm h$ reads
\begin{equation}
    w^h(\bm r) = -\hbar\gamma\bm h\cdot\left[\bm S^{(1)}(\bm r) + \bm S^{(2)}(\bm r)\right],
\end{equation}
where $\gamma 
= 2\mu_B / \hbar (s / S)$ (where $s \approx 3.56/2$~\cite{Scheie:2022aa}) is the effective gyromagnetic ratio. The Larmor frequency $\bm\Omega_i$ in the layer $i$ can be expressed through
\begin{multline}
    {\hbar}\bm \Omega_i = \A\sum_{n'}J_{FM}(n')\bm S^{(i)}(n') - J_{int}\A \bm S^{(\bar i)} \\ - 2\A\left(K_xS^{(i)}_x\bm e_x + K_zS^{(i)}_z\bm e_z\right) + {\hbar}\gamma\bm h,
\end{multline}
where $\bar i$ corresponds to the opposite layer.

The equilibrium spin projections are $\bm S^{(1)}_0 = S_0\bm e_y$ and $\bm S^{(2)}_0 = -S_0\bm e_y$. We introduce the two-dimensional layer magnetization (per unit area) $\bm m_i = \hbar\gamma\left(\bm S^{(i)} - \bm S^{(i)}_0\right) \propto\bm h$. Spin dynamics equations are
\begin{equation}
    \frac{d\bm S^{(i)}}{dt} + \bm S^{(i)}\times \bm \Omega_i = 0.
\end{equation}
Using the plane wave expansion $\bm m_i = \bm m_ie^{{\rm i}\bm k\cdot\bm r - {\rm i}\omega t}$, where $\bm k = (k_x, k_y)$ is the in-plane wavevector, the dynamics equations are
\begin{subequations}
    \begin{multline}
        -{\rm i}\hbar\omega \bm m_1 = \A S_0\left[J_{FM}(\bm k) - J_{FM}(0)\right]\bm m_1\times \bm e_y \\ -J_{int}\A S_0(\bm m_1 + \bm m_2)\times \bm e_y \\ - 2\A S_0\left(K_xm_{1x}\bm e_x + K_zm_{2z}\bm e_z\right)\times \bm e_y + \hbar^2\gamma^2S_0\bm h\times \bm e_y,
    \end{multline}
    \begin{multline}
        -{\rm i}\hbar\omega \bm m_2 = - \A S_0\left[J_{FM}(\bm k) - J_{FM}(0)\right]\bm m_2\times \bm e_y \\ J_{int}\A S_0(\bm m_1 + \bm m_2)\times \bm e_y \\ + 2\A S_0\left(K_xm_{2x}\bm e_x + K_zm_{2z}\bm e_z\right)\times \bm e_y + \hbar^2\gamma^2S_0\bm h\times \bm e_y.
    \end{multline}
\end{subequations}

Next, we notice, that $\A S_0 = S = 3/2$ is the spin of one cite. We introduce the total magnetization $\bm m = \bm m_1 + \bm m_2$ and antiferromagnetic vector $\bm l = \bm m_1 - \bm m_2$
\begin{subequations}
    \begin{equation}
        \label{eq:l}
        \frac{{\rm i}\hbar\omega \bm m}{2S} = \left[\frac{J_{FM}(0) - J_{FM}(\bm k)}{2} + K_xl_x\bm e_x + K_zl_z\bm e_z\right]\times \bm e_y,
    \end{equation}
    \begin{multline}
        \label{eq:m}
        \frac{{\rm i}\hbar\omega \bm l}{2S} = \left[\frac{J_{FM}(0) - J_{FM}(k)}{2}\bm m + J_{int}\bm m \right. \\ \left. + K_xm_x\bm e_x + K_zm_z\bm z - \frac{\hbar^2\gamma^2}{\A}\bm h\right]\times \bm e_y.
    \end{multline}
\end{subequations}
The antiferromagnetic vector can be expressed via magnetization from Eq.~\eqref{eq:l} as
\begin{subequations}
    \label{eqs:l}
    \begin{equation}
        l_x = \frac{{\rm i}\hbar\omega}{2S\left[K_z + \frac{J_{FM}(0) - J_{FM}(\bm k)}{2}\right]}m_z,
    \end{equation}
    \begin{equation}
        l_z = \frac{{\rm i}\hbar\omega}{2S\left[K_x + \frac{J_{FM}(0) - J_{FM}(\bm k)}{2}\right]}m_x.
    \end{equation}
\end{subequations}
After substitution of Eqs.~\eqref{eqs:l} in Eq.~\eqref{eq:m} we get
\begin{subequations}
    \begin{multline}
        \left[\frac{J_{FM}(0) - J_{FM}(\bm k)}{2} + J_{int} + K_z\right]m_z \\ - \frac{\hbar^2\omega^2}{4S^2\left[\frac{J_{FM}(0) - J_{FM}(\bm k)}{2} + K_x\right]}m_z = \frac{\hbar^2\gamma^2}{\A}h_z,
    \end{multline}
    \begin{multline}
        \left[\frac{J_{FM}(0) - J_{FM}(\bm k)}{2} + J_{int} + K_x\right]m_x \\ - \frac{\hbar^2\omega^2}{4S^2\left[\frac{J_{FM}(0) - J_{FM}(\bm k)}{2} + K_z\right]}m_x = \frac{\hbar^2\gamma^2}{\A}h_x.
    \end{multline}
\end{subequations}

Finally, we obtain the non-zero components of the two-dimensional magnetic susceptibility tensor, $m_\alpha = \chi^{2D}_{\alpha\beta}h_\beta$, in the form
\begin{subequations}
    \begin{multline}
        \chi^{2D}_{xx} = \frac{F_{xx}}{\Omega_x^2 - \omega^2} = \frac{4S^2\gamma^2}{\A}\left[\frac{J_{FM}(0) - J_{FM}(\bm k)}{2} + K_z\right]\\ \times\left\{\frac{4S^2}{\hbar^2}\left[\frac{J_{FM}(0) - J_{FM}(\bm k)}{2} + K_z\right]\right. \\ \left. \times \left[\frac{J_{FM}(0) - J_{FM}(\bm k)}{2} + J_{int} + K_x\right] - \omega^2\right\}^{-1},
    \end{multline}
    \begin{multline}
        \chi^{2D}_{zz} = \frac{F_{zz}}{\Omega_z^2 - \omega^2} = \frac{4S^2\gamma^2}{\A}\left[\frac{J_{FM}(0) - J_{FM}(\bm k)}{2} + K_x\right]\\\times \left\{\frac{4S^2}{\hbar^2}\left[\frac{J_{FM}(0) - J_{FM}(\bm k)}{2} + K_x\right]\right. \\\times \left. \left[\frac{J_{FM}(0) - J_{FM}(\bm k)}{2} + J_{int} + K_z\right] - \omega^2\right\}^{-1}.
    \end{multline}
\end{subequations}
The poles of susceptibility provide the energies of magnon modes: $E^+_{\bm k}$ follows from the pole of $\chi^{2D}_{zz}$ and $E^-_{\bm k}$ follows from the pole of $\chi^{2D}_{xx}$ in agreement with the main text.

\section{Finite thickness effects}
\label{app:FiniteThickness}

Let us now take into account the effects of finite thickness of multilayer samples where the number of layers is sufficiently large to allow for the semiclassical approach. Note that intralayer interaction terms in the Hamiltonian~\eqref{eq:w(r)} do not change in multilayers. However, the interlayer antiferromagnetic interaction term $w^{AFM}(\bm r)$ doubles up due to the fact that two neighboring layers surround a given one. We study the magnetostatic regime (that corresponds to a non-retarded dipole-dipole interaction studied in the main text for a bilayer) where the magnetic field is solenoidal,
\begin{equation}
    \rot \bm h = 0,
\end{equation}
that gives us the parametrization via a scalar function $\theta = \theta(z)e^{{\rm i}\bm k\cdot\bm r}$, with
\begin{equation}
    \bm h = \grad \theta = {\rm i}\bm k\theta + \frac{d\theta(z)}{dz}\bm e_z.
\end{equation}
The equation for flux reads
\begin{multline}
    0 = \divv \bm b = \divv(\bm h + 4\pi \bm m) = -(1 + 4\pi\chi_{xx})k_x^2\theta(z) \\ - k_y^2\theta(z) + \frac{d}{dz}\left[\left(1 + 4\pi\chi_{zz}\right)\frac{d\theta(z)}{dz}\right].
\end{multline}
Hereafter $\bm m$ is the bulk magnetization (magnetic moment per unit volume) and we use the bulk susceptibility, $\chi_{\alpha\beta} = \chi^{2D}_{\alpha\beta}/d$, where $d$ is the effective thickness of bilayer. Also $J_{int}$ may be renormalized by a factor of $\sim 2$ due to the interaction not with one, but with a couple of neighboring layers.

We consider the material with finite thickness $L$ in vacuum. The boundary conditions are
\begin{subequations}
    \begin{equation}
        \label{bc:tau}
        \bm h_\tau = i\bm k\theta = {\rm const},
    \end{equation}
    \begin{equation}
        \label{bc:n}
        b_n = (1 + 4\pi \chi_{zz})\frac{d\theta(z)}{dz} = {\rm const}.
    \end{equation}
\end{subequations}
The Eq.~\eqref{bc:tau} gives the continuity of $\theta(z)$, while the Eq.~\eqref{bc:n} gives the relation for the dervatives of $\theta(z)$. 
In vacuum, $|z|>L/2$ magnetostatic equation is
\begin{equation}
    \frac{d^2\theta(z)}{dz^2} - k^2\theta(z) = 0,
\end{equation}
that leads to the exponential solutions $\propto \exp(\pm kz)$, deacreasing at $\pm \infty$.

Inside the magnetic layer we have
\begin{equation}
    (1 + 4\pi\chi_{zz})\frac{d^2\theta(z)}{dz^2} - (k^2 + 4\pi\chi_{xx}k_x^2)\theta(z) = 0,
\end{equation}
that transforms in the oscillation equation
\begin{equation}
    \frac{d^2\theta(z)}{dz^2} + \Tilde{k}^2\theta(z) = 0,
\end{equation}
where the wavevector is
\begin{equation}
    \Tilde{k}^2 = -\frac{(1 + 4\pi\chi_{xx})k_x^2 + k_y^2}{1 + 4\pi\chi_{zz}}.
\end{equation}
We consider two types of modes: even and odd.

\subsection{Even mode}
We look for the solution in a form
\begin{subequations}
    \begin{equation}
        \theta(z) = A\cos{\Tilde{k}z}, \quad |z| < L/2,
    \end{equation}
    \begin{equation}
        \theta(z) = Be^{-k(|z| - L / 2)}.
    \end{equation}
\end{subequations}
The boundary condition~\eqref{bc:tau} gives
\begin{equation}
    B = A\cos{\frac{\Tilde{k}L}{2}},
\end{equation}
while the boundary contition~\eqref{bc:n} gives
\begin{equation}
    -(1 + 4\pi\chi_{zz})\Tilde{k}A\sin{\frac{\Tilde{k}L}{2}} = -kB.
\end{equation}
The dispersion relation is
\begin{equation}
    k = (1 + 4\pi\chi_{zz})\Tilde{k}\tan{\frac{\Tilde{k}L}{2}}.
\end{equation}
For the thin films $kL\ll 1$
\begin{equation}
    k = (1 + 4\pi\chi_{zz})\frac{\Tilde{k}^2L}{2} = -\left[(1 + 4\pi\chi_{xx})k_x^2 + k_y^2\right]\frac{L}{2}.
\end{equation}
Thus, we find the magnon energy
\begin{multline}
\label{positive:disp:macro}
    \hbar^2\omega^2 = \Omega_x^2 + \frac{2\pi F_{xx}k_x^2{N}}{k} \\ = 4S^2\left[\frac{J_{FM}(0) - J_{FM}(\bm k)}{2} + K_z\right]\\\times \left[\frac{J_{FM}(0) - J_{FM}(\bm k)}{2} + J_{int} + K_x\right] \\ + \frac{8\pi S^2\hbar^2\gamma^2}{\A}\left[\frac{J_{FM}(0) - J_{FM}(\bm k)}{2} + K_z\right]\frac{k_x^2{N}}{k},
\end{multline}
where $N = L/d$ is the number of bilayers in the sample.

\subsection{Odd mode}
\begin{figure}[ht]
    \centering
    \includegraphics[width=\linewidth]{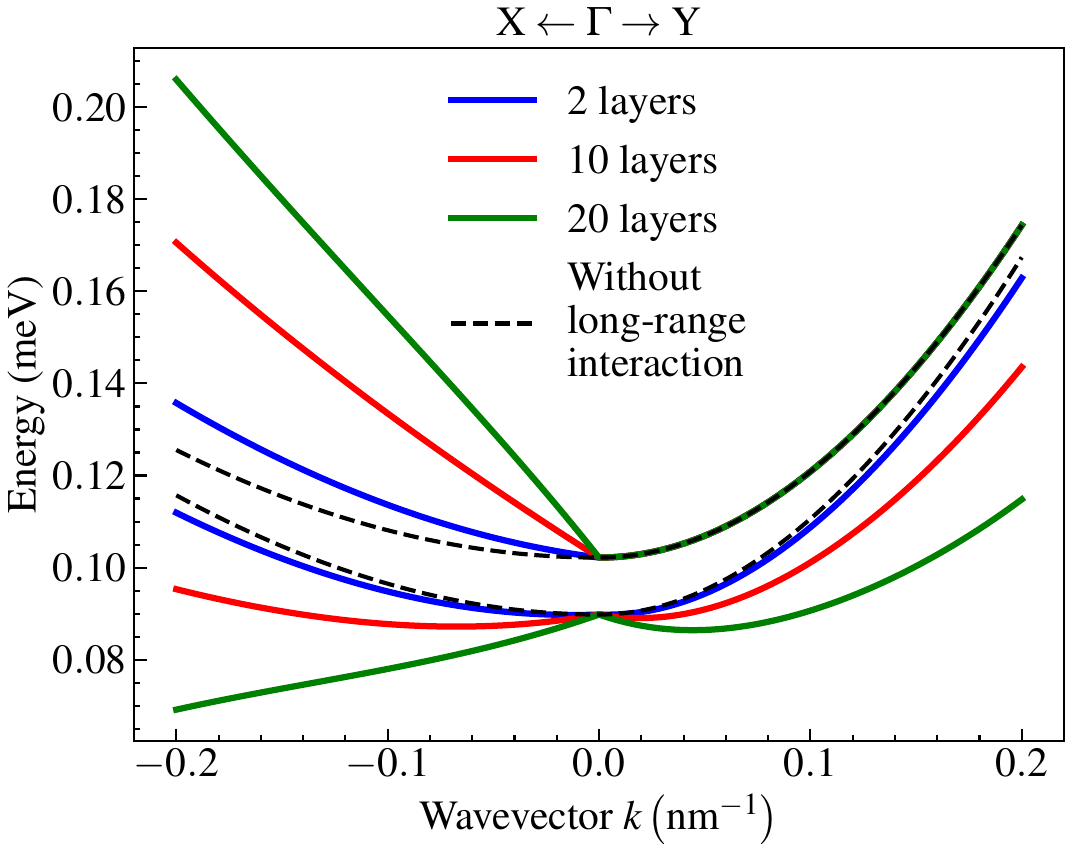}
    \caption{Magnon dispersion at small wavevectors~$k$ for different number of layers. Group velocity reaches $u_+\approx -200$~m/s for $20$~layers.}
    \label{fig:disper:dd}
\end{figure}

We look for the solution in form
\begin{subequations}
    \begin{equation}
        \theta(z) = A\sin{\Tilde{k}z}, \quad |z| < L/2,
    \end{equation}
    \begin{equation}
        \theta(z) = {\rm sign}(z)Be^{-k(|z| - L / 2)}.
    \end{equation}
\end{subequations}
The boundary condition~\eqref{bc:tau} gives
\begin{equation}
    B = A\sin{\frac{\Tilde{k}L}{2}},
\end{equation}
while the boundary condition~\eqref{bc:n} gives
\begin{equation}
    (1 + 4\pi\chi_{zz})\Tilde{k}A\cos{\frac{\Tilde{k}L}{2}} = -kB.
\end{equation}
The dispersion relation is
\begin{equation}
    k\tan{\frac{\Tilde{k}L}{2}} = -(1 + 4\pi\chi_{zz})\Tilde{k}.
\end{equation}
For the thin films $kL\ll 1$, that follows $\Tilde{k}L \sim \sqrt{kL} \ll 1$,
\begin{equation}
    1 + 4\pi\chi_{zz} = -\frac{kL}{2}.
\end{equation}
Thus, we find the magnon energy
\begin{equation}
    \hbar^2\omega^2 = \Omega_z^2 + \frac{4\pi F_{zz}/d}{1 + \frac{kL}{2}} = \Omega_z^2 + 4\pi F_{zz}/d - 2\pi F_{zz}kN.
\end{equation}
The term $4\pi F_{zz}$ is the depolarization (Lamb) shift, that should be included in the ``short-range'' parameters of the magnetic system. Finally, the magnon energy is
\begin{multline}
\label{negative:disp:macro}
    \hbar^2\omega^2 = \Omega_z^2 - 2\pi F_{zz}kN \\ = 4S^2\left[\frac{J_{FM}(0) - J_{FM}(\bm k)}{2} + K_x\right]\\\times \left[\frac{J_{FM}(0) - J_{FM}(\bm k)}{2} + J_{int} + K_z\right] \\ - \frac{8\pi S^2\hbar^2\gamma^2}{\A}\left[\frac{J_{FM}(0) - J_{FM}(\bm k)}{2} + K_x\right]kN.
\end{multline}
The $k$-linear corrections to the magnon dispersion result from the long-range dipole-dipole interactions between Cr spins. The effective velocities increase linearly with the number of layers, Fig.~\ref{fig:disper:dd}.

Noteworthy, in the limit of $k\to 0$ the decomposition of Eqs.~\eqref{positive:disp:macro} and \eqref{negative:disp:macro} yields, at $N=1$, expressions derived in Sec.~\ref{sec:dd} of the main text for a bilayer.

\section{\addZakhar{Transformation coefficients $X^{\pm\pm}$}}
\label{app:Xs}

It is instructive to analyze the values of the transformation coefficients $X^{\pm\pm}_{\bm k}$ in Eq.~\eqref{X:coeff} that relate the magnon operators $e_{\bm k}, f_{\bm k}, \ldots$ with the operators $a_{\bm k}, b_{\bm k}, \ldots$ in Eq.~\eqref{transition:fin}. For sufficiently large momenta $\bm k$ where the role of anisotropy, dipole-dipole interactions, and antiferromagnetic interlayer coupling becomes minor the magnon dispersion is controlled by the intralayer ferromagnetic interaction with the result 
\begin{equation}
    E_{\bm k}^\pm =S [J^{FM}(0) - J^{FM}(\bm k)] = \sum_{\alpha = x, y}\hbar^2k_\alpha^2/(2m_m^\alpha),
\end{equation}
where $m_m^\alpha$ are magnon effective masses along the principal axes of the structure, and the wavevector is assumed to be small compared to the Brillouin zone size. The coefficients at sufficiently large wavevectors $\bm k$ tends to
\begin{equation}
\label{transition:fin:asympt:0}
    X^{\pm+}_{\boldsymbol k} \to \sqrt{2}, \qquad X^{\pm-}_{\boldsymbol k} \to 0.
\end{equation}
These simplified asymptotic expressions can be readily derived noting that for large enough $\bm k$ magnons in the first and second layer become independent. In this case, the modes are degenerate and equation~\eqref{transition:fin} describes the basis change from intralayer ``circularly'' polarized magnon $a_{\bm k}, b_{\bm k}$ to the $z$- and $x$-polarized combinations $e_{\bm k}$ and $f_{\bm k}$. 

\begin{figure}[ht]
    \centering
    \includegraphics[width=\linewidth]{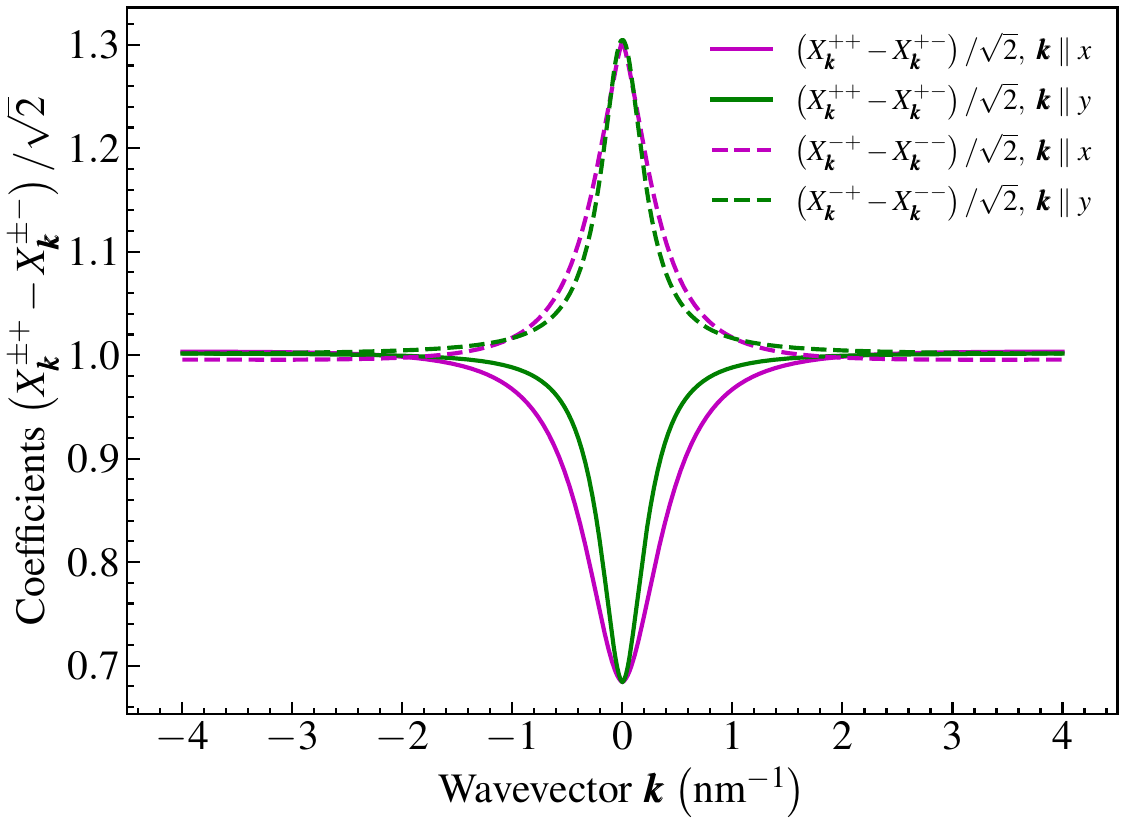}
    \caption{Relevant combinations of transformation coefficients for both branches as a function of wavevector.} 
    \label{fig:ME}
\end{figure}

On opposite, at small values of magnon momentum $\bm k$ important in Sec.~\ref{sec:transport} the values of these coefficients are different:
\begin{equation}
    \label{eq:X0ratio}
    \frac{X_0^{-+} - X_0^{--}}{X^{++}_0 - X^{+-}_0} = \left[\frac{K_z(J_{int} + K_z)}{K_x(J_{int} + K_x)}\right]^{1/4} = 1.9.
\end{equation}
The results of numerical calculation of the coefficient showing both limits and the transition between them at characteristic wavevector $|\bm k|\sim 1$~nm$^{-1}$ are presented in Fig.~\ref{fig:ME}.

\section{Exciton-magnon polaron binding energy}
\label{Appendix:XMP}
For numerical estimations we take the wavefunction of both intralayer and interlayer excitons in a form
\begin{equation}
    \varphi(\boldsymbol \rho) = \frac1{\sqrt{2\pi r_xr_y}}\exp\left(-\frac{\rho_x^2}{4r_x^2}-\frac{\rho_y^2}{4r_y^2}\right),
\end{equation}
where following~\cite{Semina:2025} we take $r_x = 0.6$~nm and $r_y = 1.8$~nm, root mean square distances between electron and hole (anisotropic Bohr radii). Exciton-magnon polaron has nonzero binding energy at zero temperature $E_{XMP}^0$ due to the virtual magnon creation and annihilation processes. The energy increases with the increase of the temperature $E_{XMP}(T)$ due to the thermal magnon population, Eq.~\eqref{thermal:magnon}.

The simple analytical approximation neglecting the dispersion [$\Delta(0, \bm q) \equiv \Delta$] gives exciton-magnon polaron energy at zero temperature as
\begin{equation}
    E_{XMP}(0) = \frac{\mathcal{A}_0}{\pi \Tilde{r}_x\Tilde{r}_y}\frac{t_e^2}{16S\Delta},
\end{equation}
where $\Tilde{r}_\alpha = m_h^\alpha r_\alpha / M^\alpha$ is characteristic size of inverse wavevector in Eq.~\eqref{Feq}. The prefactor has clear physical meaning: exciton-magnon polaron binding energy is proportional to the ratio of the unit cell to the exciton area. The polaron binding energy at elevated temperature ($k_BT \gg E_0^M \equiv E^\pm_{\bm k=0} \approx 0.1~\text{meV}$, the magnon cut-off energy) grows linearly with temperature
\begin{multline}
    \label{eq:num1}
    E_{XMP}(T) = \frac{t_e^2}{4\pi S\Delta}\frac{\sqrt{m_m^xm_m^y}\mathcal{A}_0}{\hbar^2}k_BT\\ \times \left[\ln\left(\frac{2\hbar^2}{m_m^y\Tilde{r}_y^2E_0^\pm}\right) - \gamma\right].
\end{multline}
where $\gamma \approx 0.577$ is Euler's constant. In the above approximations we took into account the quasi-one-dimensional nature of CrSBr
\[
E^\pm_0,~\frac{\hbar^2}{2m_m^y\Tilde{r}_y^2} 
\ll \frac{\hbar^2}{2m_m^x\Tilde{r}_x^2}.
\]

The better, but more complicated approximation ($E_{XMP}'$), taking into account the interlayer exciton dispersion, $\Delta(0, \bm q)$, gives small corrections both to the energy at zero temperature
\begin{equation}
    E_{XMP}'(0) = \zeta\left(\frac{\Delta}{E^X_x}\right)E_{XMP}(0),
\end{equation}
and to the $T$-linear term
\begin{equation}
    \label{eq:num2}
    E_{XMP}'(T) = \zeta\left(\frac{\Delta}{E^X_y}\right)E_{XMP}(T) - 2\frac{m_m^x}{M_X}E_{XMP}'(0)\frac{k_BT}{\Delta},
\end{equation}
where we introduced the function
\begin{equation}
    \zeta(x) = \sqrt{\pi x}\exp(x)\text{Erfc}(x),
\end{equation}
$E^X_\alpha = \hbar^2/(2M^\alpha\Tilde{r}_\alpha^2)$ is the kinetic energy term with wavevector along $\alpha\in\{x, y\}$ direction at the limit wavevector $q_\alpha = \Tilde{r}_\alpha^{-1}$.

Note that the expressions and numerical results may differ a bit for different relations between $|t_e|$ and $|t_h|$.

\section{\addZakhar{Exciton-magnon interaction: technical details}}
\label{app:technical:x-m}

In the second order perturbation theory the mixing element between states $\Psi^{(1)}_\uparrow$ and $\Psi^{(2)}_\downarrow$ is
\begin{equation}
    \label{eq:H41}
    \mathcal{H}_{e,i}^{41} = \frac{\mathcal{H}_{e,i}^{43}\mathcal{H}_{e,i}^{31}}{-SE_S} + \frac{\mathcal{H}_{e,i}^{42}\mathcal{H}_{e,i}^{21}}{-SE_S} = \frac{t_e(b^\dagger_i + a_i)}{\sqrt{2S}},
\end{equation}
where as a superscripts we use Hamiltonian $\mathcal{H}_{e,i}$ matrix row and column, and $-SE_S$ in denominator is the energy difference between the ground and opposite spin states.

The matrix elements of the exciton-magnon interaction with allowance for both the electron and hole tunneling read:
\begin{subequations}
    \label{HXM-momentum2:me}
\begin{multline}
    \mathcal{H}_{XM}(\bm k_2, \bm k_1)_{11} = \\ -\frac{1}{\mathcal{N}}\sum_{\bm k'}{c^\dagger_{\bm k' - \bm k_2}}\frac{t_{\bm k_2 - \bm k'}^{e*} t^e_{\bm k' - \bm k_1}+t^{h*}_{\bm k_2 - \bm k'}t^h_{\bm k' - \bm k_1}}{2S\Delta}{c_{\bm k' - \bm k_1}},
\end{multline}
\begin{multline}
    \mathcal{H}_{XM}(\bm k_2, \bm k_1)_{12} = \\ -\frac{1}{\mathcal{N}}\sum_{\bm k'}{c^\dagger_{\bm k' - \bm k_2}}\frac{t^{h*}_{\bm k_2 - \bm k'}t^e_{\bm k' - \bm k_1} + t^{e*}_{\bm k_2 - \bm k'}t^h_{\bm k' - \bm k_1}}{2S\Delta}{c^\dagger_{\bm k_1 - \bm k'}},
\end{multline}
\begin{multline}
    \mathcal{H}_{XM}(\bm k_2, \bm k_1)_{21} = \\ -\frac{1}{\mathcal{N}}\sum_{\bm k'}{c_{\bm k_2 - \bm k'}}\frac{t^{h*}_{\bm k_2 - \bm k'}t^e_{\bm k' - \bm k_1} + t^{e*}_{\bm k_2 - \bm k'}t^h_{\bm k' - \bm k_1}}{2S\Delta}{c_{\bm k' - \bm k_1}},
\end{multline}
\begin{multline}
    \mathcal{H}_{XM}(\bm k_2, \bm k_1)_{22} = \\ -\frac{1}{\mathcal{N}}\sum_{\bm k'}{c_{\bm k_2 - \bm k'}}\frac{t^{e*}_{\bm k_2 - \bm k'}t^e_{\bm k' - \bm k_1}+t^{h*}_{\bm k_2 - \bm k'}t^h_{\bm k' - \bm k_1}}{2S\Delta}{c^\dagger_{\bm k_1 - \bm k'}}.
\end{multline}
\end{subequations}

\section{Details on derivation of collision integrals}
\label{app:collision}

\subsection{Scattering}\label{app:collision:sc}

To derive the contribution of the exciton-magnon scattering processes we take into account the fact that the effective masses of magnons are much larger than the effective masses of excitons and use the quasielastic approximation. 
The collision integral is
\begin{multline}
    \label{collision:sc1}
    Q^{\sigma}_{sc}\left\{g^X,g^{\sigma}\right\} = \frac{2\pi}{\hbar}|2V^{\sigma}|^2\\\times \frac{1}{\mathcal{S}^2}\sum_{\bm p, \bm q}\delta\left(E^X_{\bm k} + E^{\sigma}_{\bm p} - E^X_{\bm k + \bm q} - E^{\sigma}_{\bm p - \bm q}\right)\\\times\left\{g^X(\bm k + \bm q)\left[g_0^{\sigma}(\bm p) + 1\right]\delta g^{\sigma}(\bm p - \bm q) \right. \\ + g^X(\bm k + \bm q)\delta g^{\sigma}(\bm p)g_0^{\sigma}(\bm p - \bm q) \\ - g^X(\bm k)\delta g^{\sigma}(\bm p)\left[g_0^{\sigma}(\bm p - \bm q) + 1\right] \\ \left. - g^X(\bm k)g_0^{\sigma}(\bm p)\delta g^{\sigma}(\bm p - \bm q)\right\}.
\end{multline}
Next, we consider the same temperature of excitons and magnons, thus, the principle of detailed balance gives
\begin{multline}
    \delta\left(E^X_{\bm k} + E^{\sigma}_{\bm p} - E^X_{\bm k + \bm q} - E^{\sigma}_{\bm p - \bm q}\right) \\\times g^X(\bm k + \bm q)\left[g^{\sigma}_0(\bm p) + 1\right]g^{\sigma}_0(\bm p - \bm q) \\ = \delta\left(E^X_{\bm k} + E^{\sigma}_{\bm p} - E^X_{\bm k + \bm q} - E^{\sigma}_{\bm p - \bm q}\right) \\\times g^X(\bm k)g^{\sigma}_0(\bm p)\left[g^{\sigma}_0(\bm p - \bm q) + 1\right].
\end{multline}

The collision integral~\eqref{collision:sc1} transforms into
\begin{multline}
    Q^{\sigma}_{sc}\left\{g^X,g^{\sigma}\right\} = \frac{2\pi}{\hbar}|2V^{\sigma}|^2\\\times \frac{1}{\mathcal{S}^2}\sum_{\bm p, \bm q}\delta(E_{\bm k}^X + E_{\bm p}^{\sigma} - E_{\bm k + \bm q}^X - E_{\bm p - \bm q}^{\sigma}) \\\times g^X(\bm k) g_0^{\sigma}(\bm p)\left[g_0^{\sigma}(\bm p - \bm q) + 1\right] \\\times \left\{\left[g_0^{\sigma}(\bm p - \bm q) + 1\right]\frac{\hbar(\bm p - \bm q)\bm u}{k_BT} + g_0^{\sigma}(\bm p)\frac{\hbar\bm p\bm u}{k_BT} \right. \\ \left. - \left[g_0^{\sigma}(\bm p) + 1\right]\frac{\hbar\bm p\bm u}{k_BT} - g_0^{\sigma}(\bm p - \bm q)\frac{\hbar(\bm p - \bm q)\bm u}{k_BT}\right\} \\ = -\frac{2\pi}{\hbar}|2V^{\sigma}|^2\frac{1}{\mathcal{S}^2}\sum_{\bm p, \bm q}\delta(E_{\bm k}^X + E_{\bm p}^{\sigma} - E_{\bm k + \bm q}^X - E_{\bm p - \bm q}^{\sigma}) \\ \times g^X(\bm k) g_0^{\sigma}(\bm p)\left[g_0^{\sigma}(\bm p - \bm q) + 1\right]\frac{\hbar\bm q \bm u}{k_BT}.
\end{multline}

\addZakhar{The magnon distibution function $g_0^\sigma(\bm p)$ is singular for $E^\sigma_{\bm p} \ll k_BT$ at $\bm p \to 0$ (the condition $E^\sigma_0 \ll k_B T$ is held for all the experimental temperature range. Under this assumption $g_0^\sigma \gg 1$ and the unity can be neglected in the brackets. Within the quasielastic approximation} we can rewrite the collision integral in a form
\begin{multline}
    Q^{\sigma}_{sc}\left\{g^X,g^{\sigma}\right\} = -\frac{2\pi}{\hbar}|2V^{\sigma}|^2g^X(\bm k)\\\times \frac{1}{\mathcal{S}^2}\sum_{\bm q}\delta(E^X_{\bm k} - E^X_{\bm k + \bm q})\frac{\hbar \bm q\bm u}{k_BT}\sum_{\bm p}g_0^{\sigma}(\bm p)g_0^{\sigma}(\bm p - \bm q).
\end{multline}

The internal sum can be taken analytically
\begin{equation}
    \frac1{\mathcal{S}}\sum_{\bm p}g_0^{\sigma}(\bm p)g_0^\sigma(\bm p - \bm q) = \mathcal D^M\frac{(k_BT)^2}{E_0^{\sigma}}\addZakhar{\xi(\bm q)},
\end{equation}
where $\mathcal{D}^M = \sqrt{m_{m}^{x}m_{m}^{y}}/(2\pi\hbar^2)$ is two-dimensional magnon density of states and temperature is much greater, than the magnon cut-off frequency $k_BT \gg E_0^{\sigma} \sim 0.1$~meV. \addZakhar{The form-factor $\xi(\bm q)$ is due to the separation of $\bm p$ and $\bm p - \bm q$ in reciprocal space
\begin{multline}
    \xi(\bm q) = \frac{\ln\left[1+2\sqrt{\epsilon^\sigma_{\bm q/2} - 1}\left(\sqrt{\epsilon^\sigma_{\bm q/2}} + \sqrt{\epsilon^\sigma_{\bm q/2} - 1}\right)\right]}{2\sqrt{\epsilon^\sigma_{\bm q / 2}}\sqrt{\epsilon^\sigma_{\bm q / 2} - 1}} \\ = \begin{cases}
        1, & \epsilon^\sigma_{\bm q} \ll 1, \\
        2\frac{\ln\epsilon^\sigma_{\bm q}}{\epsilon^\sigma_{\bm q}}, & 1 \ll \epsilon^\sigma_{\bm q} \ll \frac{k_BT}{E^\sigma_0}.
    \end{cases}
\end{multline}
where $\epsilon^\sigma(\bm q) = E^\sigma_{\bm q} / E^\sigma_0$. In elastic processes $\bm q \lesssim \bm k$ is of the order of thermal exciton wavevector. As the magnon mass is much higher than exciton mass, $E^\sigma_{\bm q} \ll k_BT$ for all the processes.
}

\begin{figure}[ht]
    \centering
    \includegraphics[width=\linewidth]{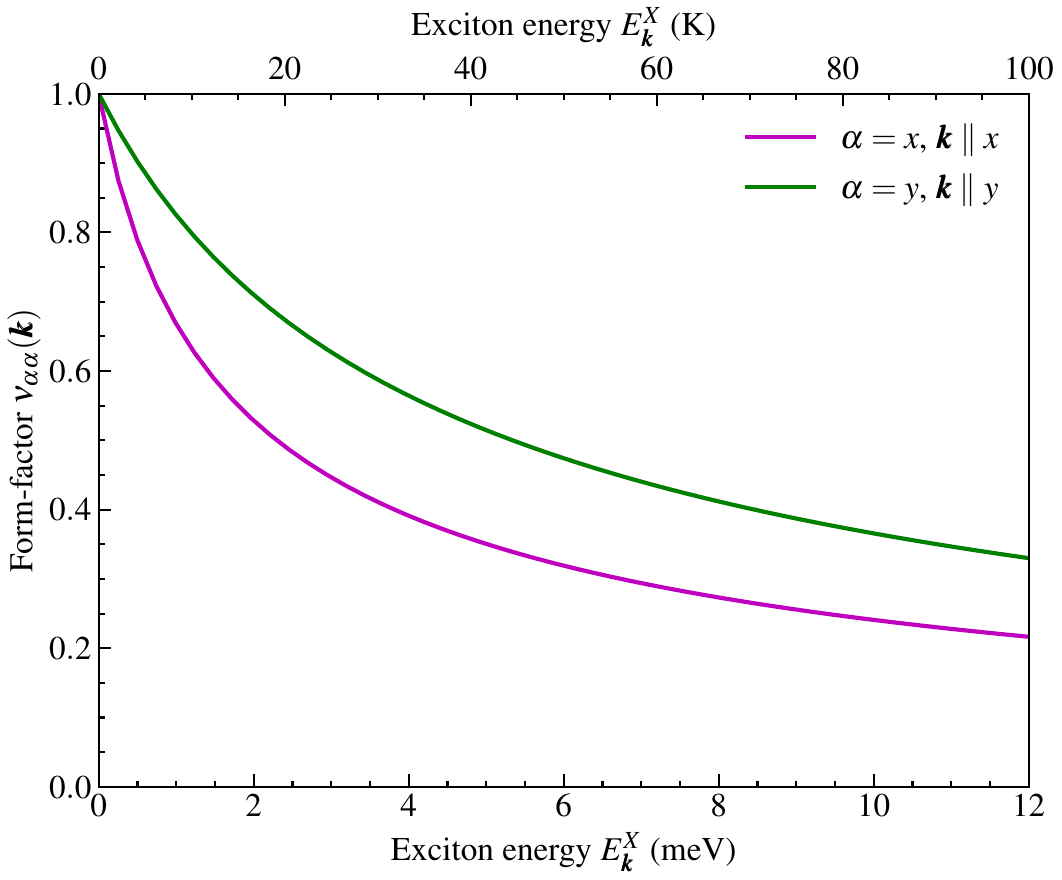}
    \caption{\addZakhar{The form-factor $\nu_{\alpha\beta}\left(E_{\bm k}^X\right)$ as a function of exciton energy (in energy scale at the bottom and in temperature scale at the top axis), Eq.~\eqref{eq:nu}. Magenta line is $\nu_{xx}(\bm k)$ for $\bm k$ along $x$-axis while green line is $\nu_{yy}(\bm k)$ for $\bm k$ along $y$-axis. For the wavevectors $\bm k$ along the Cartesian axis off-diagonal matrix components are zero. $E_0^\sigma = 0.1$~meV.}}
    \label{fig:nu}
\end{figure}

The drift term can also be calculated easily
\begin{equation}
    \label{eq:nu}
    -\frac1{\mathcal{S}}\sum_{\bm q}\delta(E^X_{\bm k} - E^X_{\bm k + \bm q})\frac{\hbar \bm q\bm u}{k_BT}\xi(\bm q) = \hbar\mathcal{D}^X\frac{\addZakhar{\nu_{\alpha\beta}\left(\bm k\right)}\addZakhar{k_\alpha u_{\beta}}}{k_BT},
\end{equation}
where $\mathcal{D}^X = \sqrt{M^{x}M^{y}}/(2\pi\hbar^2)$ is two-dimensional exciton density of states\addZakhar{, $\alpha$, $\beta = \{x, y\}$ are Cartesian indices and $\nu_{\alpha\beta}(\bm k)$ is the form-factor. Despite the anisotropic dispersion of excitons and magnons the form-factor is almost isotropic and tends to the Kronecker's $\delta$-symbol at $\bm k \to 0$: $\nu_{\alpha\beta}(\bm k) \to \delta_{\alpha\beta}$. In particular, this limit works under condition \eqref{condition:iso}. The form-factor $\nu_{\alpha\beta}$ energy dependence and anisotropy is demonstrated in Fig.~\ref{fig:nu}. Finally, the collision integral takes the form
\begin{equation}
    Q^{\sigma}_{sc}\left\{g^X,g^{\sigma}\right\} = 8\pi|V^{\sigma}|^2\mathcal{D}^M\mathcal{D}^X \frac{k_BT}{E_0^{\sigma}}g^X(\bm k)\nu_{\alpha\beta}(\bm k)k_\alpha u_\beta.
\end{equation}
and at small exciton energies it becomes isotropic}
\begin{equation}
    \label{Q:scatt:1}
    Q^{\sigma}_{sc}\left\{g^X,g^{\sigma}\right\} = 8\pi|V^{\sigma}|^2\mathcal{D}^M\mathcal{D}^X \frac{k_BT}{E_0^{\sigma}}g^X(\bm k)\bm k\bm u.
\end{equation}
\addZakhar{In the presence of the exciton flow with the distribution given by Eq.~\eqref{x:distrib} we have instead of Eq.~\eqref{Q:scatt:1}
\begin{equation}
    Q^{\sigma}_{sc}\left\{g^X,g^{\sigma}\right\} = \addZakhar{8}\pi|V^{\sigma}|^2\mathcal{D}^M\mathcal{D}^X \frac{k_BT}{E_0^{\sigma}}g^X(\bm k)\bm k(\bm u - \bm v).
\end{equation}
}

\subsection{Two-magnon absorption}
\label{app:abs}

The collision integral in the same approximations gives
\begin{multline}
    Q^{\sigma}_{abs}\left\{g^X,g^{\sigma}\right\} = \frac{2\pi}{\hbar}|V^{\sigma}|^2\\ \times\frac{1}{\mathcal{S}^2}\sum_{\bm p, \bm q}\delta(E^X_{\bm k} + E^{\sigma}_{\bm p} + E^{\sigma}_{-\bm p + \bm q} - E^X_{\bm k + \bm q})\\\times\left\{g^X(\bm k + \bm q)\left[g_0^{\sigma}(\bm p) + 1\right]\delta g^{\sigma}(-\bm p + \bm q) \right. \\ + g^X(\bm k + \bm q)\delta g^{\sigma}(\bm p)\left[g^{\sigma}_0(-\bm p + \bm q) + 1\right] \\ - g^X(\bm k)g^{\sigma}_0(\bm p)\delta g^{\sigma}(-\bm p + \bm q) \\ \left. - g^X(\bm k)\delta g^{\sigma}(\bm p)g_0(-\bm p + \bm q)\right\}
    \\ =-\frac{2\pi}{\hbar}|V^\sigma|^2\frac{g^X(\bm k)}{\mathcal{S}^2}\sum_{\bm p, q}\delta(E^X_{\bm k} + E^\sigma_{\bm p} + E^\sigma_{-\bm p + \bm q} - E^X_{\bm k + \bm q})\\ \times g^\sigma_0(\bm p)g^\sigma_0(-\bm p + \bm q)\frac{\hbar\bm q\bm u}{k_BT}.
\end{multline}
Similar transformations, taking into account, that magnons are heavier, than electrons, gives $E^\sigma_{\bm p} \sim E^\sigma_{-\bm p + \bm q} \ll E^X_{\bm k} \sim E^X_{\bm k + \bm q} \sim k_BT$ and the collision integral
\begin{multline}
    Q^{\sigma}_{abs}\left\{g^X,g^{\sigma}\right\} = -\frac{2\pi}{\hbar}|V^{\sigma}|^2g^X(\bm k)\\\times\frac{1}{\mathcal{S}^2}\sum_{\bm q}\delta(E^X_{\bm k} - E^X_{\bm k + \bm q})\sum_{\bm p}g_0^{\sigma}(\bm p)g_0^{\sigma}(-\bm p + \bm q) \frac{\hbar\bm q\bm u}{k_BT}.
\end{multline}
Using the same calculations we get, the collision integral similar to the previous one
\begin{equation}
\label{Q:abs:1}
    Q^{\sigma}_{abs}\left\{g^X,g^{\sigma}\right\} = 2\pi|V^{\sigma}|^2\mathcal{D}^M\mathcal{D}^X \frac{k_BT}{E_0^{\sigma}}g^X(\bm k)\addZakhar{\nu_{\alpha\beta}(\bm k)k_\alpha u_\beta}.
\end{equation}
Similarly, in the case of exciton distribution given by Eq.~\eqref{x:distrib} we have instead of Eq.~\eqref{Q:abs:1} \addZakhar{in isotropic limit}
\begin{equation}
    Q^{\sigma}_{abs}\left\{g^X,g^{\sigma}\right\} = 2\pi|V^{\sigma}|^2\mathcal{D}^M\mathcal{D}^X\frac{k_BT}{E_0^{\sigma}}g^X_0(\bm k)\bm k(\bm u - \bm v).
\end{equation}

\subsection{Two-magnon emission}
\label{app:em}

The collision integral in the same approximations gives
\begin{multline}
    Q^{\sigma}_{em}\left\{g^X,g^{\sigma}\right\} = \frac{2\pi}{\hbar}|V^{\sigma}|^2\\ \times \frac{1}{\mathcal{S}^2}\sum_{\bm p, \bm q}\delta(E^X_{\bm k} - E^{\sigma}_{\bm p} - E^{\sigma}_{\bm p - \bm q} - E^X_{\bm k + \bm q})\\\times\left\{g^X(\bm k + \bm q)g_0^{\sigma}(-\bm p)\delta g^{\sigma}(\bm p - \bm q)\right. \\ + g^X(\bm k + \bm q)\delta g^{\sigma}(-\bm p)g^{\sigma}_0(\bm p - \bm q) \\ - g^X(\bm k)\left[g^{\sigma}_0(-\bm p) + 1\right]\delta g^{\sigma}(\bm p - \bm q) \\ \left. - g^X(\bm k)\delta g^{\sigma}(-\bm p)\left[g_0(\bm p - \bm q) + 1\right]\right\}
    \\ =-\frac{2\pi}{\hbar}|V^\sigma|^2\frac{g^X(\bm k)}{\mathcal{S}^2}\sum_{\bm p, q}\delta(E^X_{\bm k} - E^\sigma_{-\bm p} - E^\sigma_{\bm p - \bm q} - E^X_{\bm k + \bm q})\\ \times \left[g^\sigma_0(-\bm p) + 1\right]\left[g^\sigma_0(\bm p - \bm q) + 1\right]\frac{\hbar\bm q\bm u}{k_BT}.
\end{multline}
Similar transformation, taking into account, that magnons are heavier, than electrons, gives 
\begin{multline}
    Q^{\sigma}_{em}\left\{g^X,g^{\sigma}\right\} = -\frac{2\pi}{\hbar}|V^{\sigma}|^2g^X(\bm k)\\ \times\frac{1}{\mathcal{S}^2}\sum_{\bm q}\delta(E^X_{\bm k} - E^X_{\bm k + \bm q})\sum_{\bm p}g_0^{\sigma}(\bm p - \bm q)g_0^{\sigma}(-\bm p)\frac{\hbar\bm q\bm u}{k_BT}.
\end{multline}
where we again consider $E^X_{\bm k} \gg E^\sigma_0$.
Finally, the collision integral \addZakhar{in these limits is the same as for two-magnon emission
\begin{equation}
    Q^\sigma_{em}\left\{g^X, g^\sigma\right\} = Q^\sigma_{abs}\left\{g^X, g^\sigma\right\}.
\end{equation}
}

\end{document}